\documentclass[11pt,letterpaper]{article}
\usepackage{graphicx, epsfig} 
\usepackage{amsmath,amssymb,amsfonts,dsfont,mathrsfs,amsthm,mathtools}
\usepackage{gensymb, amsthm, thmtools, etoolbox, tensor}
\usepackage{bm} 
\usepackage[utf8]{inputenc}
\usepackage{cancel}
\usepackage[normalem]{ulem}
\usepackage{soul}
\usepackage{jheppub}
\usepackage{color}
\usepackage[usenames]{xcolor}
\usepackage{hyperref}
\usepackage{siunitx}
\hypersetup{linktocpage,colorlinks=true,urlcolor=blue,linkcolor=magenta,citecolor=blue}
\definecolor{blue-violet}{rgb}{0.54, 0.17, 0.89}
\definecolor{PineGreen}{cmyk}{0.92, 0, 0.59, 0.25}
\definecolor{Gray}{cmyk}{0, 0, 0, 0.50}

\newcommand{\LL}{\mathcal{L}}
\newcommand{\OO}{\mathcal{O}}
\newcommand{\RR}{\mathcal{R}}
\newcommand{\EE}{\mathcal{E}}
\newcommand{\WW}{\mathcal{W}}

\newcommand{\RRzero}{{\mathcal{R}^{(0)}}}
\newcommand{\nablazero}{\tensor{\tilde{\nabla}}{^{(0)}}}

\newcommand{\Tr}{\text{Tr}}

\begin{document}

\title{Universal renormalization procedure for higher curvature gravities in $D \leq 5$}

\author[a]{Ignacio J. Araya,}
\author[b,c]{Jos\'e D. Edelstein,}
\author[b,c]{Alberto Rivadulla S\'anchez,}
\author[b,c]{David V\'azquez Rodr\'\i guez,}
\author[b,c]{Alejandro Vilar L\'opez}

\emailAdd{ignaraya@unap.cl}
\emailAdd{jose.edelstein@usc.es}
\emailAdd{alberto.rivadulla.sanchez@usc.es}
\emailAdd{davidvazquez.rodriguez@usc.es}
\emailAdd{alejandrovilar.lopez@usc.es}

\affiliation[a]{Instituto de Ciencias Exactas y Naturales, Facultad de Ciencias, Universidad Arturo Prat, Avenida Arturo Prat Chac\'on 2120, 1110939, Iquique, Chile}
\affiliation[b]{Departamento de F\'\i sica de Part\'\i culas, Universidade de Santiago de Compostela, E-15782 Santiago de Compostela, Spain}
\affiliation[c]{Instituto Galego de F\'\i sica de Altas Enerx\'\i as (IGFAE), Universidade de Santiago de Compostela, E-15782 Santiago de Compostela, Spain}

\abstract{We implement a universal method for renormalizing AdS gravity actions applicable to arbitrary higher curvature theories in up to five dimensions. The renormalization procedure considers the extrinsic counterterm for Einstein-AdS gravity given by the \textit{Kounterterms} scheme, but with a theory-dependent coupling constant that is fixed by the requirement of renormalization for the vacuum solution. This method is shown to work for a generic higher curvature gravity with arbitrary couplings except for a zero measure subset, which includes well-known examples where the asymptotic behavior is modified and the AdS vacua are degenerate, such as Chern-Simons gravity in 5D, Conformal Gravity in 4D and New Massive Gravity in 3D. In order to show the universality of the scheme, we perform a decomposition of the equations of motion into their normal and tangential components with respect to the Poincare coordinate and study the Fefferman-Graham expansion of the metric. We verify the cancellation of divergences of the on-shell action and the well-posedness of the variational principle.}

\maketitle

\section{Introduction}

Higher curvature gravity (HCG) has been extensively studied as a possible effective field theory extension of Einstein's general relativity (GR), motivated by the fact that curvature terms of arbitrarily high order are compatible with the diffeomorphism invariance symmetry, and that these terms can generically be considered as quantum corrections to the Einstein-Hilbert action at high energies \cite{Utiyama:1962sn}. The presence of higher curvature terms on the effective gravitational action are a robust prediction of string theory \cite{Gross:1986iv} ---there are indeed hints of a possible mutual implication \cite{Camanho:2014apa}, and the resulting effective theories have better UV properties than GR. However, generic higher curvature theories have dynamical instabilities and ghost-like degrees of freedom. The best studied HCG is Lovelock gravity, which was constructed in order to have second-order field equations and no dynamical (Ostrogradsky) instabilities \cite{Lovelock:1971yv}. Other HCGs that have been constructed to be of second order around specific backgrounds, such as spherically-symmetric or cosmological (FLRW) backgrounds, are quasi-topological gravities \cite{Oliva:2010eb} and generalized quasi-topological gravities (GQGs) \cite{Bueno:2019ltp,Bueno:2019ycr,Arciniega:2018tnn}. Stability conditions have been studied perturbatively for these theories and other concerns regarding their dynamics have been already addressed in the literature.

In the context of the gauge/gravity duality \cite{Maldacena:1997re}, the role of HCGs can be understood as a tool to accommodate the description of a broader class of conformal field theories (CFTs), having for example different central charges and anomaly coefficients than those dual to GR \cite{Nojiri:1999mh}. The holographic properties of HCGs such as Lovelock and GQGs have also been discussed in the literature, where anomaly coefficients and central charges have been holographically computed and restrictions have been placed on the couplings for the higher curvature terms based on CFT considerations, such as the unitarity requirement for the CFT dual to the particular HCG in the bulk (see \cite{deBoer:2009pn,Camanho:2009vw,Buchel:2009sk} for Lanczos-Gauss-Bonnet, \cite{deBoer:2009gx,Camanho:2009hu,Camanho:2013pda} for Lovelock and \cite{Myers:2010jv,Parvizi:2017boc} for GQGs).

As it is standard in the saddle-point approximation of anti-de Sitter/CFT (AdS/CFT) holography, the use of the on-shell classical gravity action as the generating functional for connected correlators of the dual CFT requires the finiteness of said action and also requires the corresponding variational principle to be well-posed. The latter is needed in order for arbitrary variations of the action to correspond entirely to variations with respect to the boundary field values, which are identified as the CFT sources. In the case of GR and Lovelock gravity, both requirements have been implemented following the holographic renormalization (HR) prescription (\cite{deHaro:2000vlm,Henningson:1999xi,Skenderis:2002wp,Emparan:1999pm,Balasubramanian:1999re} for GR and \cite{Yale:2011dq} for Lovelock). In its original proposal, this prescription considers first fixing the variational principle at an arbitrary radial boundary in the bulk of an asymptotically locally AdS (AlAdS) manifold, using the standard Gibbons-Hawking-York (GHY) term or the Myers term \cite{Myers:1987yn} respectively. Then, the radius at which the boundary is placed is taken as a regulator of the theory and it is extended to the conformal boundary of the space. This produces divergent terms at different orders in the regulator, which can be thus isolated. Finally, boundary terms which are intrinsic in the induced metric at the boundary, and thus depend on its Riemannian curvature and covariant derivatives thereof, are added in order to remove said divergences without modifying the variational principle. 

In the case of arbitrary HCGs, the boundary term required for fixing the variational principle at fixed radius and the HR counterterms required for the finiteness of the action on AlAdS spaces are not known. However, it has been suggested that for Einsteinian cubic gravity (a GQG of cubic order in $D=4$), the action can be rendered finite with the same combination of the GHY and the HR counterterms used in GR, up to a coupling-dependent overall factor \cite{Bueno:2018xqc}. It is therefore suggestive to explore the possibility of defining a universal renormalization prescription, that would work for an arbitrary HCG, allowing not only to cancel the divergences in the gravitational action, but also to pose the variational principle properly. 

In this work, we develop such a generic renormalization scheme for arbitrary HCGs, in spacetimes with dimension $D\leq5$. In order to do so, we study the radial decomposition of the equations of motion (EOM) for the arbitrary HCG, expanded in the Poincare coordinate $z$ considering the Fefferman-Graham (FG) expansion of the metric \cite{AST_1985__S131__95_0}, which is standard for AlAdS manifolds. We show that the divergent terms which contribute to the on-shell action at the boundary (for bulk $D\leq5$) depend only on the first four FG coefficients of the metric (associated to the powers in $z$ up to $z^3$). We also show that for the generic combination of couplings, the second coefficient (of the order $z$ term) and the fourth coefficient (order $z^3$) are zero, and that the third coefficient (order $z^2$) has a universal form. Using these facts, we are able to show that the extrinsic boundary counterterms discussed in \cite{Olea:2005gb,Olea:2006vd}, with theory-dependent coupling constants, successfully implement the renormalization and fix the variational principle at the conformal boundary. In showing this, we consider the asymptotic equivalence between the $\textit{Kounterterm}$ renormalization scheme and the HR prescription for manifolds of up to $D=5$, as discussed in \cite{Anastasiou:2020zwc}.

The paper is organized as follows. In section \ref{section2}, we consider the radially decomposed EOM of an arbitrary HCG of the form $\mathcal{L}\left(\text{Riemann}\right)$, in order to check that for the generic case, the second and fourth coefficients in the FG expansion of the metric ($g^{(1)}_{ij}$ and $g^{(3)}_{ij}$) are zero, which is needed in order to verify the renormalization of the theory. In section \ref{section3}, we present the extrinsic boundary counterterms, with the correct theory-dependent couplings, which implement the renormalization in arbitrary HCGs (in $D\leq5$ spacetime dimensions). In section \ref{section4}, we verify the cancellation of divergences in HCG actions renormalized with the generic prescription. In section \ref{section5}, we verify that the addition of the extrinsic counterterm used in the generic renormalization prescription achieves the well-posedness of the variational principle for a generic HCG. Finally, in section \ref{section6}, we summarize and discuss our results.

\section{Projected equations of motion in HCG}\label{section2}

We consider a HCG in $D\leq5$ whose action is given by
\begin{equation}
	I = \int_M d^D\!X \sqrt{-G} \, \mathcal{L}\left( R_{\mu\nu}^{\rho\sigma} \right) ~,
\end{equation}
where the Lagrangian includes any possible term constructed from arbitrary contractions of the Riemann tensor and the metric. The equations of motion are given by \cite{Padmanabhan:2011ex}
\begin{equation}
	\mathcal{E}^\nu_\mu = P_{\mu\alpha}^{\beta\gamma} R^{\nu\alpha}_{\beta\gamma} - \frac{1}{2} \delta^\nu_\mu \mathcal{L} - 2 \nabla^\alpha \nabla_\beta P_{\mu\alpha}^{\beta \nu} ~,
	\label{eq Expression equations of motion most general Padmanabhan}
\end{equation}
where the tensor $P^{\mu\nu}_{\rho\sigma}$ is defined as
\begin{equation}
	P^{\mu\nu}_{\rho\sigma} = \frac{\partial \mathcal{L}}{\partial R^{\rho\sigma}_{\mu\nu}} ~.
	\label{P-tensor}
\end{equation}
Since we are interested in renormalizing theories of gravity on AlAdS backgrounds, we consider that the Riemann tensor near the AdS boundary behaves as
\begin{equation}
	{R}^{\rho\sigma}_{\mu\nu} \rightarrow \, -\frac{1}{L^{2}} ( \delta^{\rho}_{\mu} \delta^{\sigma}_{\nu} - \delta^{\rho}_{\nu} \delta^{\sigma}_{\mu}) ~,
	\label{eq Asymptotically AdS Riemann tensor}
\end{equation}
where $L$ is the effective AdS radius, which is related to that appearing in the cosmological term, $\Lambda = -(D-1)(D-2)/(2 L_0^2)$, through a relation $L = L(L_0)$ that depends on the particular theory. Notice that $P^{\mu\nu}_{\rho\sigma}$ has the same symmetries as the Riemann tensor \eqref{P-tensor}; thereby, close to the boundary it becomes
\begin{equation}
	P^{\mu\nu}_{\rho\sigma} \, \rightarrow \, \frac{1}{2} C(L) (\delta^\mu_\rho \delta^\nu_\sigma - \delta^\mu_\sigma \delta^\nu_\rho) ~.
	\label{eq Tensor P zeroth order general}
\end{equation}
The constant $C(L)$ can be obtained by replacing these expressions in the field equations \cite{Bueno:2016ypa},
\begin{equation}
	C(L) = - \frac{L^2}{2 (D-1)} \mathcal{L} |_\text{AdS} = \frac{L^3}{2 D(D-1)} \frac{\partial \mathcal{L} |_\text{AdS}}{\partial L} ~,
	\label{eq Constant C(L) general}
\end{equation}
where $\mathcal{L} |_\text{AdS}$ is the Lagrangian evaluated in the AdS vacuum solution. In the case of Einstein-Hilbert gravity, this constant is nothing but $1/(16 \pi G_{\text{N}})$.

Now, we consider the FG expansion of the bulk metric by splitting the coordinates, $X^\mu = (x^i, z)$, into normal and tangent to the boundary,
\begin{equation}
	ds^2 = G_{\mu\nu}\, dX^\mu dX^\nu = \frac{L^2}{z^2} dz^2 + h_{ij} (x, z) dx^i dx^j ~,
	\label{eq Metric global manifold FG expansion}
\end{equation}
where $h_{ij}$ is given by\footnote{Logarithmic terms also appear in this expansion for odd-dimensions, even in Einstein's gravity.}
\begin{equation}
	h_{ij} (x, z) = \frac{1}{z^2} g_{ij}(x, z) = \frac{1}{z^2} \left( g^{(0)}_{ij}(x) + z g^{(1)}_{ij}(x) + z^2 g^{(2)}_{ij}(x) + z^3 g^{(3)}_{ij}(x) + \cdots \right) ~, 
	\label{eq Metric tangent manifold FG expansion}
\end{equation}
$z$ being the Poincare coordinate, defined such that the asymptotic conformal boundary is located at $z \to 0$. The coefficients $g^{(n)}_{ij}(x)$ with $n < D-1$ are determined completely in terms of $g^{(0)}_{ij}(x)$ by means of the projected equations of motion $\EE^z_z = 0$, $\EE^i_j = 0$ and $\EE^z_i = 0$. Since we are considering $D\leq5$, we want to scrutinize the coefficients in this expansion up to order $z^3$, in particular to see whether they are the same as those in Einstein's gravity,
\begin{eqnarray}
	g^{(1)}_{ij} & = & 0 ~, \nonumber \\ [0.4em]
	g^{(2)}_{ij} & = & - \frac{L^2}{D-3} \left( \RR_{ij}(g^{(0)}) - \frac{1}{2(D-2)} \RR(g^{(0)})\, g^{(0)}_{ij} \right) ~,
	\label{eq Coefficients FG expansion Einstein gravity}
	\\ [0.4em]
	g^{(3)}_{ij} & = & 0 \nonumber ~,
\end{eqnarray}
or not. Here, $\RR_{ij}(g^{(0)})$ and $\RR(g^{(0)})$ are respectively the Ricci tensor and curvature scalar computed from $g^{(0)}_{ij}$. In the form \eqref{eq Metric global manifold FG expansion}, the metric is naturally decomposed in the parts that are normal ($zz$ component) and tangent to the boundary ($ij$ components). The tangent indices are raised with the inverse tangent metric $h^{ij} (x, z) = z^2 g^{ij} (x, z)$, which has an expansion in $z$ such that $h_{ik} h^{jk} = \delta^j_i$. The vector normal to the boundary, $n$, is defined as
\begin{equation}
	n = n^\mu \partial_\mu = n^z \partial_z = - \frac{z}{L} \partial_z ~, \qquad n^\mu n_\mu = 1 ~,
	\label{eq Normal vector to the boundary}
\end{equation}
where the minus sign appears due to the fact that the boundary is located at the lowest limit of the possible range of values of the radial coordinate $z$. We can also define a covariant derivative compatible with the tangent part of the metric, $\tilde{\nabla}_i$, such that
\begin{equation}
	\tilde{\nabla}_i h_{jk} = \tilde{\nabla}_i g_{jk} = 0 ~,
\end{equation}
since $h_{jk} = g_{jk}/z^2$ and $\tilde{\nabla}_i z = 0$ by definition. Since $h_{ij}$ and $g_{ij}$ depend on $z$, the tangent covariant derivative will also admit an expansion in this normal coordinate. Apart from this, the usual covariant derivative is compatible with the global metric, $\nabla_\mu G_{\rho\sigma} = 0$. For completeness, we write the explicit form of the different non-vanishing components of the Christoffel symbols computed from the metric \eqref{eq Metric global manifold FG expansion},
\begin{equation}
	\Gamma^z_{zz} (\nabla) = - \frac{1}{z} ~, \qquad \Gamma^i_{zj} (\nabla) = \frac{1}{2} h^{ik} \partial_z h_{jk} ~, \qquad   \Gamma^z_{ij} (\nabla) = - \frac{1}{2} \frac{z^2}{L^2} \partial_z h_{ij} ~,
	\label{eq Christoffel symbols explicit form}
\end{equation}
whereas $\Gamma^k_{ij} (\nabla) = \Gamma^k_{ij} (\tilde{\nabla}) + \OO(z)$. Again, since $h_{ij} = h_{ij} (x, z)$, most of these admit an expansion in powers of $z$, in terms of the different coefficients in \eqref{eq Metric tangent manifold FG expansion}. We further need to know the form of the extrinsic curvature of this metric, given by
\begin{equation}
	K_{ij} = \frac{1}{2} \LL_n h_{ij} = \frac{1}{2} n^z \partial_z h_{ij} = - \frac{1}{2} \frac{z}{L} \partial_z \left( \frac{1}{z^2} g_{ij} \right) ~.
	\label{eq Extrinsic curvature definition}
\end{equation}
Now, we can use the Gauss-Codazzi equations, together with the expression of the extrinsic curvature, to obtain the different components of the Riemann tensor that will be needed to compute the projected equations of motion,
\begin{eqnarray}
	R_{i z j z} & = & \frac{L^2}{z^2} \left( - \LL_n K_{ij} + K_{ik} \tensor{K}{^k_j} - a_i a_j + \tilde{\nabla}_{(i} a_{j)} \right) ~, \nonumber \\ [0.4em]
	R_{ijk z} & = & - n_z \left( \tilde{\nabla}_i K_{jk} - \tilde{\nabla}_j K_{ik} \right) ~, 
	\label{eq Riemann of global metric from Gauss-Codazzi equations} \\ [0.6em]
	R_{ijkl} & = & \RR_{ijkl} (h) + K_{il} K_{jk} - K_{ik} K_{jl} ~, \nonumber
\end{eqnarray}
where $a_\mu = n^\alpha \nabla_\alpha n_\mu$, and $\RR_{ijkl}(h)$ is the usual Riemann tensor of the tangent metric $h_{ij}(x,z)$, which can be expanded also in powers of $z$,
\begin{equation}
	\RR_{ijkl}(h) = \frac{1}{z^2} \RR_{ijkl}(g) = \frac{1}{z^2} \left( \RR_{ijkl} (g^{(0)}) + \OO(z) \right) ~.
	\label{eq Relation Riemann tensors of h and g0}
\end{equation}
Notice that the indices of $\RR_{ijkl}(h)$ are raised with the tangent inverse metric $h^{ij}$, while those of $\RR_{ijkl}(g)$ must be raised with $g^{ij}$.

\subsection{Vanishing of $g^{(1)}_{ij}$ in a general HCG}
\label{subsec Vanishing of g1 in a general HCG}

If we compute the different components of the Riemann tensor explicitly to next-to-leading order in the holographic coordinate, we find that they match the expresesion
\begin{equation}
	R^{\mu\nu}_{\rho\sigma} = - \frac{2}{L^2} \delta^{[\mu}_\rho \delta^{\nu]}_\sigma + z \frac{2}{L^2} \delta^{[\mu}_{[\rho} \tensor{g}{^{(1)}}^{\nu]}_{\sigma]} + \OO (z^2) ~,
	\label{eq General form Riemann lowest order}
\end{equation}
where by definition only the components of ${g^{(1)}}^\nu_\sigma$ with both indices in the directions tangent to the boundary can be non-zero. Regarding $P^{\mu\nu}_{\rho\sigma}$, we can use its definition \eqref{P-tensor} and the expression of the Riemann tensor \eqref{eq General form Riemann lowest order} to constrain its tensorial form,
\begin{equation}
	P^{\mu\nu}_{\rho\sigma} = C(L) \, \delta^{[\mu}_\rho \delta^{\nu]}_\sigma + z \left( A^{(1)}(L) \, \delta^{[\mu}_{[\rho} \, \tensor{g}{^{(1)}}^{\nu]}_{\sigma]} + B^{(1)}(L) \, \delta^{[\mu}_\rho \delta^{\nu]}_\sigma \,\Tr\,g^{(1)} \right) + \OO (z^2) ~,
	\label{eq General form tensor P constants order z1}
\end{equation}
where $C(L)$ was given in \eqref{eq Constant C(L) general}, whereas $A^{(1)}(L)$ and $B^{(1)}(L)$ are scalar functions depending on the couplings of the theory and the effective AdS radius, $L$; also, $\Tr\,g^{(1)} := {g^{(1)}}^i_i$.

The Lagrangian also appears in the general expression of the equations of motion (\ref{eq Expression equations of motion most general Padmanabhan}), so we expand it symbolically to first order in $z$,
\begin{equation}
	\LL = \LL^{(0)} + P^{\rho\sigma}_{\mu\nu} \delta R^{\mu\nu}_{\rho\sigma} + \cdots = \LL^{(0)} + z \, \LL^{(1)} + \OO (z^2) ~,
	\label{eq General form Lagrangian order z1}
\end{equation}
where $\LL^{(0)}$ is the Lagrangian evaluated in the background solution \eqref{eq Asymptotically AdS Riemann tensor} and, given that $\delta R^{\mu\nu}_{\rho\sigma}$ is the deviation of the Riemann tensor with respect to this background, thereby already of order $z$,
\begin{equation}
	\LL^{(1)} = \frac{D-1}{L^2} C(L) \,\Tr\,g^{(1)} ~.
	\label{eq Coefficient L1 general}
\end{equation}
With all these ingredients, the equations of motion, decomposed into their radial and tangential components and expanded in powers of $z$, read:
\begin{align}
	\EE^z_z & = \left(\frac{(1-D) C(L)}{L^2} - \frac{\LL^{(0)}}{2}\right) + \frac{z}{2 L^2} \left(a^{(1)}(L) + (D-1)b^{(1)}(L)\right)\, \Tr\,g^{(1)} + \OO (z^2) ~, 
	\label{eq Projected equation nn general theory order z1} \\[0.4em]
	\EE^i_j & = \left( \frac{(1-D) C(L)}{L^2} - \frac{\LL^{(0)}}{2} \right) \delta^i_j + \frac{z (D-2)}{2 L^2} \left[ a^{(1)}(L)\, \tensor{g}{^{(1)}^i_j} + b^{(1)}(L)\, \delta^i_j\, \Tr\,g^{(1)} \right] + \OO(z^2) ~,
	\label{eq Projected equation pp general theory order z1} \\[0.5em]
	\EE^z_i & =
	\frac{1}{2 (D-2) L^2} \left( a^{(1)}(L)\, \tilde{\nabla}_j \tensor{g}{^{(1)}^j_i} + b^{(1)}(L)\, \tilde{\nabla}_i \Tr\,g^{(1)} \right) + \OO(z^3) ~,
	\label{eq Projected equation np general theory order z1}	
\end{align}
where
\begin{eqnarray}
	a^{(1)}(L) & := & C(L) - A^{(1)}(L) ~, \label{eq Constants in EoMs at order z1 general} \\ [0.5em] 
	b^{(1)}(L) & := & - C(L) - A^{(1)}(L) - 4 B^{(1)}(L) ~. \nonumber
	\label{g1 condition}
\end{eqnarray}
The zeroth order gives the aforementioned result for $C(L)$ and all the information of the equations at the lowest orders is encoded in $a^{(1)}(L)$ and $b^{(1)}(L)$. The main feature of these equations is that, except in some particular cases, they imply
\begin{equation}
    \tensor{g}{^{(1)}^i_j} = 0 ~.
    \label{vanishing-g1}
\end{equation}
More specifically, the equations of motion fix the $\OO(z)$ FG coefficient to be zero, except in the following cases: 
\begin{itemize}
\item If $a^{(1)}(L) \neq 0$ and $b^{(1)}(L)=-\frac{a^{(1)}(L)}{(D-1)}$, the off-diagonal components of $\tensor{g}{^{(1)}^i_j}$ are fixed to zero and the elements in the diagonal are fixed to be equal to each other, but their value is free.
\item If $a^{(1)}(L)=0$ and $b^{(1)}(L) \neq 0$, $\Tr g^{(1)}$ is fixed to zero, but otherwise it is free. 
\item And finally, if both $a^{(1)}(L)=0$ and $b^{(1)}(L) = 0$, $\tensor{g}{^{(1)}^i_j}$ is fully free and therefore it is not restricted by the equations of motion.
\end{itemize}
This is interesting as it means that, for the generic HCG theory, its equations of motion require the $\OO(z)$ coefficient in the FG expansion of the metric to vanish, just like for Einstein-AdS gravity. This universality is analogous to the universality of the $\OO(z^2)$ coefficient as implied by the PBH transformations \cite{Imbimbo:1999bj}, and both are central to the applicability of the generic renormalization procedure presented in this work.

\subsubsection*{An example: quadratic curvature gravity}

Let us illustrate this further by working out explicitly the case of quadratic curvature gravity. Its action can be written as
\begin{equation}
	I = \int_M d^D\!X \sqrt{-G} \left[ \frac{1}{\kappa} (R - 2 \Lambda_0) + \alpha_1 R_{\mu\nu} R^{\mu\nu} + \alpha_2 R^2 + \alpha_3 \chi_\text{GB} \right] ~,
	\label{eq Lagrangian quadratic curvature gravity general}
\end{equation}
where $\chi_\text{GB}$ is the Lanczos-Gauss-Bonnet combination. Once we compute and plug in the values of $C(L)$, $a^{(1)}(L)$ and $b^{(1)}(L)$, we have
\begin{align}
    C(L) & = \frac{1}{\kappa} - \frac{2 (D-1)}{L^2} \left( \alpha_1 + D \alpha_2 + \frac{(D-2)(D-3)}{D-1} \alpha_3 \right) ~, \\[0.5em]
	a^{(1)}(L) & = \frac{1}{\kappa} - \frac{1}{L^2} \bigg[ (3D - 4) \alpha_1 + 2 D (D-1) \alpha_2 + 2 (D-3) (D-4) \alpha_3 \bigg] ~, \\[0.5em]
	b^{(1)}(L) & = - \frac{1}{\kappa} + \frac{1}{L^2} \bigg[ (D-4) \alpha_1 + 2 (D-1) (D-4) \alpha_2 + 2 (D-3) (D-4) \alpha_3 \bigg] ~.
\end{align}
These equations imply $g^{(1)}_{ij} = 0$, unless the conditions discussed after \eqref{vanishing-g1} are met. Some examples of quadratic curvature gravity theories where $a^{(1)}(L) = b^{(1)}(L) = 0$, such that $g^{(1)}_{ij}$ is not fixed by the equations of motion, include:
\begin{itemize}
	\item Einstein-Lanczos-Gauss-Bonnet gravity at the (dimensionally continued) Chern-Simons point. In \cite{Banados:2004zt}, the authors find that in this theory the coefficients of the FG expasion are not fixed by the equations of motion, as they vanish identically.
	\item Conformal Gravity in 4 dimensions \cite{Anastasiou:2016jix, Anastasiou:2020mik,Grumiller:2013mxa}, where it is found again that the equations of motion at the lowest orders vanish for any form of the coefficients in the expansion.
	\item New Massive Gravity \cite{Kwon:2011jz, Cunliff:2013en} at the special point (in the language of \cite{Kwon:2011jz}).
\end{itemize}
A feature all these cases share is that their AdS vacua are degenerate. Indeed, we checked that this is true for any quadratic or cubic gravity theory fulfilling $a^{(1)}(L) = b^{(1)}(L) = 0$. While it is true that these particular cases allow for $g^{(1)}_{ij} \neq 0$, they do not enforce it. In particular, $g^{(1)}_{ij}$ cannot be given as an expression in terms of $g^{(0)}_{ij}$, thereby its value must be fixed as a boundary condition \cite{Grumiller:2013mxa}. Therefore, in general we could pick $g^{(1)}_{ij} = 0$ for any theory in vacuum, and build our discussion on top of this assumption.

We have also repeated this analysis for general theories of gravity with cubic contractions of the curvature tensors. The results are written in appendix \ref{appendix Conditions to leave g1 undetermined in cubic theories of gravity}.

\subsection{Universality of $g^{(2)}_{ij}$ from the PBH transformations}
\label{subsec Universality of g2 from the PBH transformation}

The aim of this section is to obtain the form of the coefficient $g^{(2)}_{ij}$ in terms of the conformal boundary metric $g^{(0)}_{ij}$ for general theories of gravity. While this could in principle be computed from the equations of motion $\EE^\mu_\nu = 0$ at order $z^2$, here we will review the approach of \cite{Imbimbo:1999bj}, which can be applied directly to any HCG with an asymptotically AdS solution.

The argument is based on the invariance of the bulk metric under PBH (Penrose-Brown-Henneaux) transformations, which are a subset of the bulk diffeomorphisms that reduce to Weyl transformations on the boundary metric, and leave the form of the bulk metric unchanged \cite{Imbimbo:1999bj}. Imposing this on \eqref{eq Metric global manifold FG expansion}, one obtains transformation properties for the coefficients in the FG expansion of the tangent metric $g_{ij}(x, z)$.

The authors of \cite{Imbimbo:1999bj} take the terms with odd powers of $z$ in the expansion of the tangent metric \eqref{eq Metric tangent manifold FG expansion} to be zero. This had already been shown to be true for Einstein gravity \cite{AST_1985__S131__95_0}, and they assume that it holds for a general theory with an AdS solution. In our case we have proven that $g^{(1)}_{ij} = 0$ for general HCGs, except in some very particular cases.\footnote{Strictly speaking, it has been known for a while that there are theories where $g^{(1)}_{ij}$ is nonzero \cite{Grumiller:2013mxa,Kwon:2011jz}.} Since we are restricting ourselves to this more general scenario, and we are currently interested in the coefficient $g^{(2)}_{ij}$, it is enought to take the results of \cite{Imbimbo:1999bj}. In particular, it is found that invariance under PBH transformations constrains the form of $g^{(2)}_{ij}$ to be
\begin{equation}
	g^{(2)}_{ij} = - \frac{L^2}{D-3} \left( \RR_{ij} (g^{(0)}) - \frac{1}{2 (D-2)} \RR (g^{(0)})\, g^{(0)}_{ij} \right) ~,
	\label{eq Coefficient g2 general form}
\end{equation}
where we took care of the fact that, due to scaling properties, the term $g^{(n)}_{ij}$ contains $n$ derivatives with respect to the tangent coordinates $x^i$. 

We should note that, since the argument builds upon invariance under boundary Weyl transformations, it fails to capture further contributions made of contractions of the Weyl tensor of the boundary that might appear.\footnote{Of course, ${\WW^{(0)}}^{ij}_{ik}$ vanishes identically, so this term is ruled out. However, at this order there could appear contractions, for example, with the schematic form of $\sqrt{\WW^{(0)} \WW^{(0)}}$ with two free indices, as seen for the case of Chern-Simons gravity in \cite{Banados:2004zt}.} However, for $2-$ and $3-$dimensional boundaries, which correspond respectively to $D = 3$ and $D = 4$ bulk dimensions, the Weyl tensor is identically zero, so these contributions do not occur. While for $D = 5$ this is not true, when treating the well-posedness of the variational problem we will need to assume asymptotic conformal flatness \cite{Anastasiou:2019ldc}, which implies that the Weyl tensor of the boundary must be zero. Therefore, we see that under these assumptions the coefficient $g^{(2)}_{ij}$ takes the same form in a general HCG as in Einstein's theory \eqref{eq Coefficients FG expansion Einstein gravity}.

\subsection{Vanishing of $g^{(3)}_{ij}$ in a general HCG}
\label{subsec Vanishing of g3 in a general HCG}

Until now, we have found that, for a general HCG, the coefficients $g^{(1)}_{ij}$ and $g^{(2)}_{ij}$ in the FG expansion \eqref{eq Metric tangent manifold FG expansion} take the same form as in Einstein gravity, written in \eqref{eq Coefficients FG expansion Einstein gravity}. We will show shortly that this is true also for the coefficient multiplied by $z^3$, and in the next sections we will see that these are the only relevant ones in order to study the renormalization of theories with up to 5 bulk dimensions.

Following the same logic as in section \ref{subsec Vanishing of g1 in a general HCG}, we expect the coefficient $g^{(3)}_{ij}$ to be fixed, in general, by the contributions of the equations of motion multiplied by $z^3$. In order to obtain this we need to expand the different objects that appear in these equations as written in \eqref{eq Expression equations of motion most general Padmanabhan}, and in particular we use symbolic expressions for both $\LL$ and $P^{\mu\nu}_{\rho\sigma}$. Since we have already fixed $g^{(1)}_{ij} = 0$, and we know the form of the zeroth order coefficients in these two objects through equations \eqref{eq Tensor P zeroth order general} and \eqref{eq Constant C(L) general}, their expansions can be written as
\begin{eqnarray}
	\LL & = - \displaystyle\frac{2 (D-1)}{L^2}C(L) + z^2 \LL^{(2)} + z^3 \LL^{(3)} + \cdots ~, 
	\label{eq Expansion lagrangian to 3rd order} \\ [0.7em]
	P^{\mu\nu}_{\alpha\beta} & = \displaystyle C(L)\,\delta^{[\mu}_\alpha \delta^{\nu]}_\beta + z^2 {P^{(2)}}^{\mu\nu}_{\alpha \beta} + z^3 {P^{(3)}}^{\mu\nu}_{\alpha \beta} + \cdots ~. 
	\label{eq Expansion tensor P to 3rd order}
\end{eqnarray}
We are interested only on the third order terms in the equations, so we should understand the form of the coefficient ${P^{(3)}}^{\mu\nu}_{\alpha\beta}$. This will depend on the theory, of course, but we can follow the same reasoning as before and use its tensorial structure to write the components ${P^{(3)}}^{ij}_{kl}$ and ${P^{(3)}}^{iz}_{jz}$ as the combinations
\begin{equation}
    \begin{aligned}
        {P^{(3)}}^{ij}_{kl} & = A^{(3)}(L) \delta^{[i}_{[k} {g^{(3)}}^{j]}_{l]} + B^{(3)}(L) \delta^{[i}_{[k} \delta^{j]}_{l]} \Tr\,g^{(3)} ~, \\[0.6em]
        {P^{(3)}}^{iz}_{jz} & = D^{(3)}(L) {g^{(3)}}^i_j + E^{(3)}(L) \delta^i_j \Tr\,g^{(3)} ~, 
    \end{aligned}
    \label{eq Expansion tensor P 3rd order in terms of g3}
\end{equation}
where the constants $A^{(3)}(L)$, $B^{(3)}(L)$, $D^{(3)}(L)$ and $E^{(3)}(L)$ depend on the AdS radius $L$ and the higher-curvature couplings. Since $g^{(1)}_{ij} = 0$, terms of the form $\tensor{g}{^{(1)}^i_j} \tensor{g}{^{(2)}^k_l}$ will not appear in these general expressions.  Also, the components with one index in the normal direction, such as ${P^{(3)}}^{ij}_{zk}$, will have terms proportional to $\tilde{\nabla} g^{(2)}$, but these appear in the equations of motion at higher orders in $z$, as they must be contracted with $n^z = z/L$. Regarding the expansion of the Lagrangian, $\LL^{(3)}$ cannot have contributions of the form $\tilde{\nabla}_i \tensor{g}{^{(2)}^j_k}$, as there are no objects with an odd number of tangent indices to contract it producing a term of order $z^3$. Therefore, it can only contain terms that are proportional to $\Tr\,g^{(3)}$, and if we expand it as in equation \eqref{eq General form Lagrangian order z1} we see that they can only be produced by ${P^{(0)}}^{\mu\nu}_{\rho\sigma} {\delta R^{(3)}}^{\rho\sigma}_{\mu\nu}$. Plugging in the third order components of the Riemann tensor,
\begin{equation}
    {R^{(3)}}^{ij}_{kl} = \frac{6}{L^2} \delta^{[i}_{[k} \tensor{g}{^{(3)}}^{j]}_{l]} ~, \qquad {R^{(3)}}^{z i}_{z j} = - \frac{3}{2 L^2} {g^{(3)}}^i_j ~,
\end{equation}  
we find
\begin{equation}
    \LL^{(3)} = \frac{3 (D-3)}{L^2} C(L) \,\Tr\,g^{(3)} ~.
\end{equation}
With all of this, we can compute the terms of third order in the projections of the equations of motion, which read
\begin{align}
    {\EE^{(3)}}^z_z & = \frac{1}{2 L^2} \left( a^{(3)}(L) + (D-1) b^{(3)}(L) \right) \Tr\,g^{(3)} ~, 
    \label{eq Projected equation nn general theory order z3} \\[0.5em]
    {\EE^{(3)}}^i_j & = \frac{D-4}{2 L^2} \left[ a^{(3)}(L) {g^{(3)}}^i_j + b^{(3)}(L)\delta^i_j \Tr\,g^{(3)} \right] ~,
    \label{eq Projected equation pp general theory order z3}
\end{align}
where the constants $a^{(3)}(L)$ and $b^{(3)}(L)$ are related to those introduced in \eqref{eq Expansion tensor P 3rd order in terms of g3} as
\begin{equation}
    \begin{aligned}
        a^{(3)}(L) & = 3 C(L) + 4 (D-6) D^{(3)}(L) - (D-3) A^{(3)}(L) ~, \\[0.5em]
        b^{(3)}(L) & = - 3 C(L) - A^{(3)}(L) - 2 (D-2) B^{(3)}(L) + 4 (D-6) E^{(3)} (L) ~.
    \end{aligned}   
\end{equation}
For completeness, we give the values of the constants $a^{(3)}(L)$ and $b^{(3)}(L)$ for general quadratic and cubic theories of gravity in appendix \ref{appendix Constants in the projected equations of motion at order z3 for general quadratic and cubic theories}.

From the discussion above on the different contributions to $\LL^{(3)}$ and ${P^{(3)}}^{\mu\nu}_{\alpha\beta}$, it is clear that the equation ${\EE^{(3)}}^z_z = 0$ fixes ---for general\footnote{Notice that $\Tr\,g^{(3)}$ must vanish for $D = 4$ while $\tensor{g}{^{(3)}^i_j}$ is left undetermined. This is expected given that in three dimensions $\tensor{g}{^{(3)}^i_j}$ is dual to the stress-energy tensor and there is no conformal anomaly.} HCGs--- $\Tr\,g^{(3)} = 0$, and this in turn means, when we consider the equation ${\EE^{(3)}}^i_j = 0$, that
\begin{equation}
	\tensor{g}{^{(3)}^i_j} = 0 ~.
\end{equation}
As we commented when fixing $g^{(1)}_{ij} = 0$, there are families of theories for which the value of $g^{(3)}_{ij}$ is not determined by the equations of motion. In particular, the same analysis discussed after \eqref{vanishing-g1} applies to $g^{(3)}$ as well, but considering the $a^{(3)}(L)$ and $b^{(3)}(L)$ coefficients instead. All the quadratic and cubic theories considered in section \ref{subsec Vanishing of g1 in a general HCG} and appendix \ref{appendix Conditions to leave g1 undetermined in cubic theories of gravity}, at the particular points mentioned, allow for $g^{(3)}_{ij} \neq 0$ even when choosing $g^{(1)}_{ij} = 0$ as a boundary condition.\footnote{With the exception of New Massive Gravity, since it is a $3-$dimensional theory, and therefore the coefficient $g^{(3)}_{ij}$ is sub-normalizable and should not be considered in the expansion \eqref{eq Metric tangent manifold FG expansion}.} However, the conditions that leave $g^{(3)}_{ij}$ undetermined do not imply the degeneracy of the different AdS vacua.

\section{Renormalization counterterms for generic HCG in D$\leq$5}\label{section3}

It was suggested in \cite{Bueno:2018xqc}, for Einsteinian cubic gravity, that one can renormalize the action using the same boundary term that is used in the Holographic Renormalization of Einstein-AdS gravity, {\it i.e.}, the GHY term plus the HR counterterm, but with a coupling-dependent overall coefficient. The renormalization was framed in terms of the cancellation of divergences of the gravity action, and not in terms of the well-posedness of the variational principle, as for higher-curvature gravities (with the exception of Lovelock) this is an open problem. As it is explained in what follows, this idea for generating counterterms based on the Einstein-AdS case can be generalized to arbitrary HCGs considering the asymptotic behaviour of AlAdS spaces.

When considering pure AdS vacua, a minimal requirement for the renormalization procedure is to render the Euclidean on-shell action equal to either zero or the vacuum energy of the maximally-symmetric configuration. As it is usual, the vacuum energy appears in odd-dimensional bulk manifolds, and in the context of AdS/CFT, it is related to the Casimir energy in the CFT side. One can then assume that the boundary term for HCGs is equal to the one for Einstein gravity but with a coupling-dependent overall factor. This overall factor can then be fixed by requiring the cancellation of divergences in the action for the maximally-symmetric solution. As in the case of $\mathcal{L}(\text{Riemann})$ theories, said action evaluated in the vacuum solution is proportional to the AdS volume, with an overall constant dependent of the couplings of the theory. One can then check if the same boundary term works for other AlAdS solutions besides the pure AdS configuration.

A similar approach was pursued in \cite{Giribet:2018hck} and \cite{Giribet:2020aks}, where the authors considered some counterterms with a multiplicative constant ---that matches the prescription in \cite{Bueno:2018xqc}--- in order to compute the Noether-Wald charges for quadratic curvature gravity theories in even-dimensional asymptotically AdS spacetimes. In \cite{Anastasiou:2021swo} the same terms are introduced to obtain renormalized entanglement entropies.

The counterterms considered in these last three references are, however, different from the usual HR proposal. The latter prescription produces a series of terms, whose complexity depends on the dimension and can not be expressed in any closed form. The new approach adds to the action some topological quantities dubbed \textit{Kounterterms} ---because they can be naturally written in terms of the extrinsic curvature of the boundary---, and were originally proposed in \cite{Aros:1999kt, Mora:2004kb, Mora:2004rx, Olea:2005gb, Olea:2006vd, Miskovic:2009bm} to renormalize the Einstein-Hilbert action and obtain a well-posed variational principle, which then allows to compute finite conserved charges in AdS gravity. Moreover, another interesting application of this method is the computation of renormalized entanglement entropies \cite{Anastasiou:2017xjr, Anastasiou:2018rla, Anastasiou:2019ldc}.

In the present work, we aim to expand this prescription to more general theories of gravity admitting AlAdS solutions in up to 5 dimensions, which can be expanded in terms of the radial coordinate as in \eqref{eq Metric global manifold FG expansion}, with the coefficients given in \eqref{eq Coefficients FG expansion Einstein gravity}. First of all, we will simply write the form of the Kounterterms for general even and odd dimensional bulks, as given in the literature. The only modification that we make is the multiplicative constant $C(L)$ that accompanies these boundary terms, which was defined in \eqref{eq Constant C(L) general} and is the only theory-dependent part of the entire expression. In sections \ref{section4} and \ref{section5} we will see that this constant appears naturally in the divergent terms that need to be cancelled, and hence it is motivated.

\subsection{Kounterterms for even bulk dimensions}

The Kounterterms for $D = 2n$ dimensions are given by \cite{Olea:2005gb}
\begin{equation}
	I_\text{Kt} = c_{2n-1} \int_{\partial M} d^{2n-1} x \, B_{2n-1} [h, K, \RR] ~,
	\label{eq General Kounterterms even dimensions}
\end{equation}
where $B_{2n-1}$ is the $n$-th Chern form\footnote{In these expressions, $\delta^{\mu_1 \cdots \mu_{p}}_{\nu_1 \cdots \nu_{p}}$ is the generalized Kronecker delta \cite{agacy1999generalized}, 
\begin{equation*}
	\delta^{\mu_1 \cdots \mu_{p}}_{\nu_1 \cdots \nu_{p}} = \text{det} \left[ \delta^{\mu_1}_{\nu_1} \cdots \delta^{\mu_p}_{\nu_p} \right] = p!\ \delta^{\mu_1}_{[\nu_1} \delta^{\mu_2}_{\nu_2} \cdots \delta^{\mu_p}_{\nu_p]} ~.
\end{equation*}
}
\begin{equation}
	\begin{aligned}
		B_{2n-1} & = - 2 n \sqrt{-h} \int_0^1 dt \, \delta_{j_1 \cdots j_{2n-1}}^{i_1 \cdots i_{2n-1}} K_{i_1}^{j_1} \left( \frac{1}{2} \RR^{j_2 j_3}_{i_2 i_3} - t^2 K^{j_2}_{i_2} K^{j_3}_{i_3} \right) \\
		& \hspace{2.6cm} \times \cdots \times \left( \frac{1}{2} \RR^{j_{2n-2} j_{2n-1}}_{i_{2n-2} i_{2n-1}} - t^2 K^{j_{2n-2}}_{i_{2n-2}} K^{j_{2n-1}}_{i_{2n-1}} \right) ~,
	\end{aligned}
	\label{eq Chern form general even dimensions}
\end{equation}
and we write the constant $c_{2n-1}$ as 
\begin{equation}
	c_{2n-1} = - \frac{(-L^2)^{n-1}}{n (2n-2)!} C(L) ~.
	\label{eq Coupling Kounterterm general even dimensions}
\end{equation}
This recovers the usual value of the constant for Einstein gravity, presented for example in \cite{Anastasiou:2020zwc}, since in that case $C(L) = 1 / \kappa$ with our conventions. However, we claim that this counterterm is suitable for more general theories of gravity whose Lagrangian is made of arbitrary contractions of the Riemann tensor, in particular, whose bulk is $4-$dimensional.

As shown in \cite{Anastasiou:2020zwc}, for Einstein gravity the Kounterterm \eqref{eq General Kounterterms even dimensions} is exactly equivalent to the usual HR prescription in $D = 4$ ---and, at least, in $D = 6$, as long as the boundary is conformally flat; {\it i.e.}, the Weyl tensor of the boundary vanishes---. We will see this explicitly in section \ref{section4}, when we show that it cancels the divergences of the on-shell action in 4 dimensions. Besides, this even-dimensional Kounterterm can be written also as a bulk integral, by means of Euler's theorem. In particular,
\begin{equation}
	\int_{M_{2n}} d^{2n}x \, \EE_{2n} = (4\pi)^n n! \,\chi (M_{2n}) + \int_{\partial M_{2n}} d^{2n-1}x \, B_{2n-1} ~,
	\label{eq Euler theorem}
\end{equation}
where $\chi(M_{2n})$ is the Euler characteristic of $M_{2n}$, and $\EE_{2n}$ is the $2n-$dimensional Euler density
\begin{equation}
	\EE_{2n} = \frac{\sqrt{-G}}{2^n} \delta^{\mu_1 \cdots \mu_{2n}}_{\nu_1 \cdots \nu_{2n}} R^{\nu_1\nu_2}_{\mu_1\mu_2} \cdots R^{\nu_{2n-1} \nu_{2n}}_{\mu_{2n-1}\mu_{2n}} ~.
\end{equation}
%

\subsection{Kounterterms for odd bulk dimensions}

For $D = 2n + 1$ bulk dimensions, the Kounterterm \cite{Olea:2006vd} reads: 
\begin{equation}
	I_\text{Kt} = c_{2n} \int_{\partial M} d^{2n}x \, B_{2n} [h, K, \RR] ~,
	\label{eq General Kounterterms odd dimensions}
\end{equation}
where $B_{2n}$ is given by
\begin{equation}
	\begin{aligned}
		B_{2n} & = - 2 n \sqrt{-h} \int_0^1 dt \int_0^t ds \, \delta_{j_1 \cdots j_{2n}}^{i_1 \cdots i_{2n}} K_{i_1}^{j_1} \delta_{i_2}^{j_2} \left( \frac{1}{2} \RR^{j_3 j_4}_{i_3 i_4} - t^2 K^{j_3}_{i_3} K^{j_4}_{i_4} + \frac{s^2}{L^2} \delta^{j_3}_{i_3} \delta^{j_4}_{i_4} \right) \\
		& \hspace{2cm} \times \cdots \times \left( \frac{1}{2} \RR^{j_{2n-1} j_{2n}}_{i_{2n-1} i_{2n}} - t^2 K^{j_{2n-1}}_{i_{2n-1}} K^{j_{2n}}_{i_{2n}} + \frac{s^2}{L^2} \delta^{j_{2n-1}}_{i_{2n-1}} \delta^{j_{2n}}_{i_{2n}} \right) ~,
	\end{aligned}
	\label{eq Chern form general odd dimensions}
\end{equation}
and the coupling constant $c_{2n}$ is
\begin{equation}
	c_{2n} = - \frac{(-L^2)^{n-1}}{2^{2n-2} n (n-1)!^2} C(L) ~.
	\label{eq Coupling Kounterterm general odd dimensions}
\end{equation}
As in the even-dimensional case, we recover the values of this constant presented in \cite{Anastasiou:2020zwc} if we set $C(L) = 1/\kappa$, as should be for Einstein gravity. Also, this is equivalent to the counterterms derived with the HR proposal, up to logarithmically divergent terms, in $D = 3$, $5$ and $7$, as long as the boundary is conformally flat \cite{Anastasiou:2020zwc}.

In this case, $B_{2n}$ can not be written as the pullback of a topological quantity in the $D = 2n+1$ manifold, contrasting with what is found for even dimensions. Also, the fact that $B_{2n}$ depends on the AdS radius $L$, while $B_{2n-1}$ does not, is a consequence of the topological origin of the latter.

\section{Divergence cancellation for HCG actions up to $D=5$}\label{section4}

In this section, we shall address the problem of renormalizing the action of a general HCG when evaluated on an AlAdS background. For this matter, we will first find the form of the divergent terms at the boundary, with a general expression valid for up to $D = 5$. This latter restriction comes from the fact that we are interested on holographic applications in realistic situations; {\it i.e.}, strongly coupled gauge theories in (at most) four dimensions. Then, we will analyze the divergent terms explicitly for $D = 3$, $4$ and $5$, and show that they are indeed cancelled by the Kounterterms presented in section \ref{section3}.

\subsection{Divergent terms in the on-shell action}

Let us consider the action of a general HCG in $D \leq 5$,
\begin{equation}
    I = \int_M d^D\!X \sqrt{-G} \, \mathcal{L}\left( R_{\mu\nu}^{\rho\sigma} \right) ~,
    \label{eq Action general HCG for analyzing the divergences on-shell}
\end{equation}
evaluated on an asymptotically locally AdS spacetime. In order to look at the structure of the divergences, we will write the metric as \eqref{eq Metric global manifold FG expansion}, with the coefficients of its FG expansion given by \eqref{eq Coefficients FG expansion Einstein gravity}. Since in these coordinates the boundary is located at $z \rightarrow 0$, it is very natural to identify the divergences in this region by looking at the terms with negative powers of $z$ in the expansion of the action \eqref{eq Action general HCG for analyzing the divergences on-shell}. Given the form of the metric \eqref{eq Metric global manifold FG expansion}, to leading order near $z \rightarrow 0$ the square root of the determinant behaves as
\begin{equation}
    \sqrt{-G} \sim \frac{1}{z^D} ~,
\end{equation}
and it has more contributions of higher order in $z$, which thus decay faster near the boundary. However, this is enough to identify the terms in the Lagrangian that will produce divergences. As shown in section \ref{section2}, the odd coefficients in the FG expansion up to the order that we are interested in vanish, so the on-shell Lagrangian can only have terms of the form $z^{2i}$. In the action, these produce
\begin{equation}
    \int dz \sqrt{-G} \, z^{2i} \sim \int dz \, z^{2i-D} \sim z^{2i-(D-1)} ~.
\end{equation}
So this can produce three different behaviors as $z \rightarrow 0$:
\begin{itemize}
    \item If $\displaystyle i < \frac{D-1}{2}$ the term is divergent, and it needs to be subtracted.
	\item If $\displaystyle i > \frac{D-1}{2}$ the term vanishes at the boundary.
	\item For odd spacetime dimensions, there can be contributions with $i = (D-1)/2$, which produce a logarithm divergence in the boundary. This is universal and related to the conformal anomaly of the dual CFT \cite{Anastasiou:2020zwc, Anastasiou:2019ldc}, and we will see that it is not cancelled by the topological Kounterterms.
\end{itemize}
Therefore, depending on the dimensionality of the spacetime the last term that produces divergences will be different. In our case, for up to 5 dimensions we will have to look at terms with $i \leq 2$.

In order to isolate the divergent terms, first of all we have to obtain an expansion of the Lagrangian $\LL (R^{\mu\nu}_{\rho\sigma})$ close to the boundary, which can be written as
\begin{equation}
    \begin{aligned}
        \LL & = \LL^{(0)} + P^{\mu\nu}_{\rho\sigma} \delta R^{\rho\sigma}_{\mu\nu} + \cdots \\[0.6em]
        & = \LL^{(0)} + P_{ij}^{kl} \delta R^{ij}_{kl} + 4 P^{z i}_{jk} \delta R_{z i}^{jk} + 4 P_{z i}^{z j} \delta R^{z i}_{z j} + \cdots ~,
    \end{aligned}
\end{equation}
where $\delta R^{\rho\sigma}_{\mu\nu}$ denotes the terms in the components of the Riemann tensor that are different from the background value \eqref{eq Asymptotically AdS Riemann tensor}, this is, those that depend on $z$. They can be computed using the Gauss-Codazzi equations \eqref{eq Riemann of global metric from Gauss-Codazzi equations}, finding
\begin{equation}
    \begin{aligned}
        \delta R^{ij}_{kl}& = \frac{z^2}{(D-2) (D-3)} \RRzero \delta^i_{[k} \delta^j_{l]} - \frac{4 z^2}{D-3} \RRzero^{[i}_{[k} \delta^{j]}_{l]} + z^2 \RRzero^{ij}_{kl} + \OO (z^4) ~, \\[0.6em]
        \delta R^{jk}_{z i} & = \frac{L^2 z^3}{D-3} \left( \nablazero^j \RR^k_i - \nablazero^k \RRzero^j_i \right) \\
        & \hspace{1cm} + \frac{L^2 z^3}{2 (D-3) (D-2)} \left( \delta^j_i \nablazero^k \RRzero - \delta_i^k \nablazero^j \RRzero \right) + \OO (z^5) ~, \\[0.6em]
        \delta R^{z j}_{z i} & = \OO(z^4) ~,
    \end{aligned}
\end{equation}
where we introduced the notation $\nablazero \equiv \tilde{\nabla}(g^{(0)})$ and $\RRzero \equiv \RR(g^{(0)})$ in order to write simpler expressions. Therefore, since ${P^{(0)}}^{z i}_{j k} = 0$, in the expansion of the Lagrangian it is enough to take
\begin{equation}
    \LL = \LL^{(0)} + {P^{(0)}}_{ij}^{kl} \delta R^{ij}_{kl} + \OO(z^4) ~,
\end{equation}
where $\OO(z^4)$ includes terms coming from ${P^{(0)}}^{z j}_{z i} \delta R^{z i}_{z j}$ and ${P^{(0)}}^{z i}_{j k} \delta R^{j k}_{z i}$, among other terms with higher powers of $z$ from the expansion of $P^{\mu\nu}_{\rho\sigma}$. Also, although we are not writing them explicitly, in this expression there should be terms with higher derivatives with respect to the Riemann tensor. However, since any derivative of the Lagrangian with respect to the Riemann tensor is, to the lowest possible order, constant in $z$, a term of the form $(\partial^n \LL / (\partial R)^n) (\delta R)^n$ will be at least $\OO(z^{2n})$. Therefore, second or higher derivatives are unimportant when looking for the divergences in low dimensions.

If we now compute ${P^{(0)}}_{ij}^{kl} \delta R^{ij}_{kl}$ explicitly with the expressions above for $\delta R^{\mu\nu}_{\rho\sigma}$ and ${P^{(0)}}^{\mu\nu}_{\rho\sigma}$ given by \eqref{eq Tensor P zeroth order general}, we find that it vanishes to the lowest order. So we conclude that the only divergent part of $\LL (R^{\mu\nu}_{\rho\sigma})$ is
\begin{equation}
    \LL = \LL^{(0)} + z^4 \LL^{(4)} ~,
\end{equation}
where the term $\LL^{(4)}$ will contribute only to the logarithmic divergence in $D = 5$ dimensions. It contains all the terms of order $z^4$ mentioned in the paragraph above, but its particular form is not relevant for our computations, since we assume that our method of renormalization will not cancel divergences of this type.

Now that we have expanded the Lagrangian in the coordinate $z$, we need to do the same with the determinant factor that appears in \eqref{eq Action general HCG for analyzing the divergences on-shell}. Using again the FG expansion of the metric \eqref{eq Metric global manifold FG expansion} and \eqref{eq Metric tangent manifold FG expansion}, this is
\begin{equation}
    \sqrt{-G} = \frac{L}{z} \sqrt{-h} = \frac{L \sqrt{-g^{(0)}}}{z^D} \left( 1 + \frac{z^2}{2} \Tr\,g^{(2)} + \OO(z^4) \right) ~.
    \label{DeterminantMetric}
\end{equation}
Plugging everything in, the divergent terms of the general action \eqref{eq Action general HCG for analyzing the divergences on-shell} for $D \leq 5$ are
\begin{equation}
    \begin{aligned}
        I_\text{diver} & = \int_M d^D\!X \, \sqrt{-G} \left( \LL^{(0)} + z^4 \LL^{(4)} \right) \\[0.6em]
        & = L \int d^{D-1}\!x \sqrt{-g^{(0)}} \int_{z = z_0} dz \left[ \frac{1}{z^D} \LL^{(0)} + \frac{1}{z^{D-2}} \frac12 \LL^{(0)} \Tr\,g^{(2)} + \OO(z^{4-D}) \right] ~,
    \end{aligned}
    \label{eq Divergent terms general theory general dimension}
\end{equation}
where we introduced the cutoff $z_0$ in the lower limit of integration, which must be taken to zero once the divergences are canceled. The actual form of these divergent contributions once integrated in $z$ will depend on the dimension of the spacetime, and in particular the second and third terms will produce the aforementioned logarithmic divergences at $D = 3$ and $D = 5$, respectively. However, the terms that we want to cancel, the power-law divergences, at these low dimensions always appear multiplied by $\LL^{(0)}$. This is proportional to the constant $C(L)$ appearing at the lowest order in $P^{\mu\nu}_{\rho\sigma}$ through the equation \eqref{eq Constant C(L) general}, or equivalently
\begin{equation}
    \LL^{(0)} = - \frac{2 (D-1)}{L^2} C(L) ~,
    \label{eq Relation L_0 and C(L)}
\end{equation}
Therefore, this supports our claim that the Kounterterms which cancel these divergences are the same as those introduced for Einstein gravity, with the explicit prefactor $C(L)$.

\subsection{Explicit analysis in different dimensions}

We will now show how the divergences \eqref{eq Divergent terms general theory general dimension} are cancelled by the Kounterterms introduced in section \ref{section3}, explicitly in 3-, 4- and 5-dimensional spacetimes. Notice that the computations carried out here were already done in section 3.4 of \cite{Anastasiou:2020zwc}, and the only difference in our results is the generic constant $C(L)$ multiplying both the Kounterterms and the divergent terms in the on-shell action. In order to see that, we will have to write the objects in \eqref{eq General Kounterterms even dimensions} and \eqref{eq General Kounterterms odd dimensions} in terms of the intrinsic curvature of the boundary metric $g^{(0)}_{ij}$.

In particular, provided the coefficients of the FG expansion are given by \eqref{eq Coefficients FG expansion Einstein gravity}, we will need the extrinsic curvature \eqref{eq Extrinsic curvature definition}
\begin{equation}
    \begin{aligned}
    K^i_j & = h^{ik} K_{kj} = \frac{1}{L} \delta^i_j - \frac{z^2}{L} {g^{(2)}}^i_j + \frac{z^4}{L} \left( g^{(2) ik} g^{(2)}_{jk} - 2 {g^{(4)}}^i_j \right) \\[0.6em]
    & = \frac{1}{L} \delta^i_j + z^2 \frac{L}{D-3} \left( \RRzero^i_j - \frac{1}{2(D-2)} \RRzero \delta^i_j \right) + \OO(z^4) ~,
    \end{aligned}
    \label{eq Extrinsic curvature in terms of intrinsic curvature}
\end{equation}
and the determinant of the tangent metric in terms of the intrinsic curvature \eqref{DeterminantMetric}, which by means of \eqref{eq Coefficient g2 general form} can be written as
\begin{equation}
    \sqrt{-h} = \frac{\sqrt{-g^{(0)}}}{z^{D-1}} \left( 1 - z^2 \frac{L^2}{4(D-2)} \RRzero + \OO(z^4) \right) ~.
    \label{eq Determinant tangent metric in terms of intrinsic curvature}
\end{equation}
%

\subsubsection{$D = 3$ dimensions}

In this case, the general divergent terms in \eqref{eq Divergent terms general theory general dimension} become, after integrating in $z$,
\begin{equation}
        I_\text{diver} = -C(L) \int d^2 x \, \sqrt{-g^{(0)}} \left[ \frac{2}{Lz_0^2} + L \log z_0 \RRzero \right] ~,
    \label{eq Divergent terms general theory 3 dimensions}
\end{equation}
where we used \eqref{eq Coefficients FG expansion Einstein gravity} to rewrite $\Tr\,g^{(2)}$ and \eqref{eq Relation L_0 and C(L)} for $\LL_0$. We need to find whether the Kounterterm \eqref{eq General Kounterterms odd dimensions}, particularized for $D = 2n+1 = 3$ dimensions, cancels the divergences found here. In this case, the constant $c_{2}$ and the function $B_2$ are equal to
\begin{equation}
    c_2 = - C(L) ~, \qquad B_2 = - \sqrt{-h} K ~.
\end{equation}
Therefore, replacing the determinant $h$ and the extrinsic curvature $K$ in terms of the intrinsic curvature of $g^{(0)}_{ij}$, we find that the total Kounterterm for $D = 3$ dimensions reads
\begin{equation}
    I_\text{Kt} = c_2 \int d^2x \, B_2 = C(L) \int d^2 x \sqrt{-g^{(0)}} \left[ \frac{2}{Lz_0^2} + \OO (z_0^2) \right] ~,
\end{equation}
which cancels the power-law divergence found in \eqref{eq Divergent terms general theory 3 dimensions}, but not the logarithmic one, as we had anticipated in section \ref{section3}.

\subsubsection{$D = 4$ dimensions}

The divergent terms in 4 dimensions \eqref{eq Divergent terms general theory general dimension} become
\begin{equation}
    I_\text{diver} = - C(L) \int d^3 x \, \sqrt{-g^{(0)}} \left[ \frac{2}{Lz_0^3} -  \frac{3 L}{4z_0} \RRzero \right] .
    \label{eq Divergent terms general theory 4 dimensions}
\end{equation}
The coupling of the Kounterterm that should cancel this and the second Chern form read
\begin{equation}
    c_3 = \frac{L^2}{4} C (L) ~, \qquad B_3 = - 4 \sqrt{-h} \delta^{i_1 i_2 i_3}_{j_1 j_2 j_3} K^{j_1}_{i_1} \left( \frac{1}{2} \RR^{j_2 j_3}_{i_2 i_3} - \frac{1}{3} K^{j_2}_{i_2} K^{j_3}_{i_3} \right) ~.
\end{equation}
Writing it all together, the Kounterterm for a general theory in $D = 4$ dimensions is
\begin{equation}
    I_\text{Kt} = C(L) \int d^3 x \sqrt{-g^{(0)}} \left[ \frac{2}{Lz_0^3} - \frac{3 L}{4z_0} \RRzero + \OO(z_0) \right] ~,
\end{equation}
which cancels exactly the divergent terms as written in \eqref{eq Divergent terms general theory 4 dimensions}. As was pointed out in \cite{Anastasiou:2020zwc} for the case of Einstein gravity, we find here that also for a general HCG the Kounterterm cancels exactly the divergences in AlAdS spaces in $D = 4$.

\subsubsection{$D = 5$ dimensions}

In this case, the divergent part of the action will have an additional logarithmic term, which depends on $\Tr\,g^{(4)}$ and $\LL^{(4)}$,
\begin{equation}
    I_\text{diver} = - C(L) \int d^4 x \,\sqrt{-g^{(0)}} \left[ \frac{2}{Lz_0^4} - \frac{L}{3z_0^2} \RRzero + \OO(1) \times \log z_0 \right] ~.
    \label{eq Divergent terms general theory 5 dimensions}
\end{equation}
As before, this should be regularized by the Kounterterm \eqref{eq General Kounterterms odd dimensions} particularized for $D = 2n + 1 = 5$ dimensions. The value of the constant $c_4$ and the function $B_4$ are
\begin{equation}
    c_4 = \frac{L^2}{8} C(L) ~, \qquad B_4 = - \sqrt{-h} \delta^{i_1 i_2 i_3}_{j_1 j_2 j_3} K^{j_1}_{i_1} \left( \RR^{j_2 j_3}_{i_2 i_3} - K^{j_2}_{i_2} K^{j_3}_{i_3} + \frac{1}{3 L^2} \delta^{j_2}_{i_2} \delta^{j_3}_{i_3} \right) ~.
\end{equation}
Therefore, the total Kounterterm in this case is
\begin{equation}
    I_\text{Kt} = C(L) \int d^4 x \, \sqrt{-g^{(0)}} \left[ \frac{2}{Lz_0^4} - \frac{L}{3z_0^2} \RRzero + \OO(1) \right] ~,
\end{equation}
which as in the case $D = 3$, cancels the divergences \eqref{eq Divergent terms general theory 5 dimensions} except for the logarithmic one. As mentioned before, these divergences are universal terms, proportional to the conformal anomaly of the dual field theory, and therefore they were not expected to be cancelled out by this renormalization procedure.

\section{Variational principle in HCG up to $D=5$}\label{section5}

We will now show that the same Kounterterms presented in section \ref{section3} also render finite the boundary terms that appear when varying the action, and they depend only on the variation of the metric of the conformal boundary $g^{(0)}_{ij}$, thus producing a well posed variational problem. First we will obtain the form of the boundary terms that we need to cancel for general dimensions up to 5, and then particularize the analysis to $D = 3$, 4 and 5 as done in the previous section to treat the divergences of the on-shell action.

\subsection{Divergences in the boundary term of the variation of a general HCG}

As in the preceding sections, we start by considering a general action that can contain any contraction of the Riemann tensor \eqref{eq Action general HCG for analyzing the divergences on-shell}. Its variation produces two contributions \cite{Padmanabhan:2011ex, Bueno:2016ypa},
\begin{equation}
	\delta I = \int_M d^D\! x \sqrt{-G}\, \EE_{\mu\nu} \delta G^{\mu\nu} + \epsilon \int_{\partial M} d^{D-1} \! x \, \sqrt{-h} \, n_\mu \delta v^\mu ~,
	\label{eq Variation of general action boundary and bulk terms}
\end{equation}
the first of which is proportional to the equations of motion \eqref{eq Expression equations of motion most general Padmanabhan}, and thus vanishes on-shell. The second term is a contraction of the vector normal to the boundary $\partial M$ (normalized such that $n \cdot n = \epsilon = \pm 1$) and the quantity
\begin{equation}
    \delta v^\mu = - 2 \tensor{P}{^\mu ^\rho ^\sigma _\nu} \delta \Gamma^\nu_{\rho\sigma} - 2 \nabla_\nu P^{\mu\rho\sigma\nu} \delta G_{\rho\sigma} ~.
	\label{eq Vector v mu in the boundary term of the variation of a HCG action}
\end{equation}
We want to evaluate the boundary term in a solution of the equations of motion with AdS asympotics, so we consider the metric to be given by the usual FG expansion \eqref{eq Metric global manifold FG expansion}. The vector normal to the boundary is given by \eqref{eq Normal vector to the boundary}, and therefore $\epsilon = 1$.

In order to evaluate the boundary term in $\delta I$ we need expressions for the variation of the Christoffel symbols. In particular, the ones we need can be written in terms of variations of the extrinsic curvature \eqref{eq Extrinsic curvature definition} as
\begin{equation}
    \begin{aligned}
    	\delta \Gamma^i_{jk} & = \frac{1}{2} h^{il} \left( \nabla_k \delta h_{jl} + \nabla_j \delta h_{kl} - \nabla_l \delta h_{jk} \right) ~, \\[0.4em]
    	\delta \Gamma^z_{ij} & = \frac{z}{L} \delta K_{ij} ~, \\[0.4em]
    	\delta \Gamma^i_{zj} & = - \frac{L}{z} \delta K^i_j ~.
    \end{aligned}
\end{equation}
Also, we expand the tensor $P^{\mu\nu}_{\rho\sigma}$ asymptotically as
\begin{equation}
	P^{\mu\nu}_{\rho\sigma} \equiv \frac{C(L)}{2} \left( \delta^\mu_\rho \delta^\nu_\sigma - \delta^\mu_\sigma \delta^\nu_\rho \right) + \delta P^{\mu\nu}_{\rho\sigma} ~,
\end{equation}
where $\delta P^{\mu\nu}_{\rho\sigma}$ is not a variation, but a symbolic way of writing all the terms in $P^{\mu\nu}_{\rho\sigma}$ containing powers of $z$ (see below, \eqref{eq Expansion delta P in powers of z}). As mentioned before, when evaluated on-shell the variation of the action contains only the boundary term given by (\ref{eq Vector v mu in the boundary term of the variation of a HCG action}), which after replacing the expressions above for $\delta \Gamma^\mu_{\rho\sigma}$ and $P^{\mu\nu}_{\rho\sigma}$ becomes
\begin{equation}
	\begin{aligned}
		\delta I & = - \int_{\partial M} d^{D-1}\! x \, \sqrt{-h} \, \Bigg[ C(L) \left( 2 \delta K^i_i + (h^{-1} \delta h)^i_j K_i^j \right) + 2 \delta P^{z i}_{z j} \left( 2 \delta K_i^j + (h^{-1} \delta h)^j_k K^k_i \right) \\
		& \hspace{4.5cm} - 4 n_z \nabla_l \delta P^{zi}_{jk} \, h^{jl} h^{km} \delta h_{im} - 2 n^z h^{jk} \nabla_z \delta P^{zi}_{zk} \, \delta h_{ij} \Bigg] ~,
	\end{aligned}
	\label{eq Boundary term of the variation of a general action in terms of deltaP}
\end{equation}
where $(h^{-1} \delta h)^i_j \equiv h^{ik} \delta h_{kj}$, and the variation $\delta h_{ij} = \delta g_{ij} / z^2$ can be written as a variation of the metric of the conformal boundary $\delta g^{(0)}_{ij}$ on-shell, due to the relations \eqref{eq Metric tangent manifold FG expansion} and \eqref{eq Coefficients FG expansion Einstein gravity}. Notice that if we restricted ourselves to Einstein gravity, we would have $C(L) = 1/\kappa$ and $\delta P^{\mu\nu}_{\alpha\beta} = 0$, independently of the background. Therefore, only the first term in this expression would contribute, recovering the results in appendix D of \cite{Anastasiou:2019ldc} for the boundary term of the variation on-shell.

Until now we have written this expression in terms of variations of the metric and the extrinsic curvature. However, on-shell they are related through \eqref{eq Coefficients FG expansion Einstein gravity} and \eqref{eq Extrinsic curvature definition}, and therefore the variations of $K_{ij}$ can be written as variations of the metric $g^{(0)}_{ij}$, thus leading to a well-posed Dirichlet problem once we get rid of the divergences. The next step is to expand $\delta P^{\mu\nu}_{\rho\sigma}$ in \eqref{eq Boundary term of the variation of a general action in terms of deltaP} in powers of $z$. Since we are not interested in more than $D = 5$ bulk dimensions, it is enough to keep only the terms with powers up to $z^4$, as the determinant in the integrand behaves to leading order as $\sqrt{-h} \sim z^{-(D-1)}$. Knowing that $g^{(1)}_{ij} = 0$, we can expand $\delta P^{\mu\nu}_{\rho\sigma}$ as
\begin{equation}
    \delta P^{\mu\nu}_{\rho\sigma} \equiv z^2 {P^{(2)}}^{\mu\nu}_{\rho\sigma} + z^3 {P^{(3)}}^{\mu\nu}_{\rho\sigma} + z^4 {P^{(4)}}^{\mu\nu}_{\rho\sigma} + \cdots ~.
    \label{eq Expansion delta P in powers of z}
\end{equation}
Plugging this into (\ref{eq Boundary term of the variation of a general action in terms of deltaP}) and evaluating the covariant derivatives explicitly yields
\begin{equation}
    \begin{aligned}
		\delta I & = - \int_{\partial M} d^{D-1}\! x \, \sqrt{-h} \, \Bigg[ C(L) \left( 2 \delta K^i_i + (h^{-1} \delta h)^i_j K^j_i \right) \\
		& \hspace{2.2cm} + \frac{z_0^2}{L} \left( - 2 (2D-7) {P^{(2)}}^{zk}_{zi} + 4 {P^{(2)}}^{lk}_{li} \right) (h^{-1} \delta h)^i_k \\
		& \hspace{2.3cm} + \frac{z_0^3}{L} \left( - 4(D-4) {P^{(3)}}^{zk}_{zi} + 4 {P^{(3)}}^{lk}_{li} \right) (h^{-1} \delta h)^i_k + z_0^4 \, \OO (\delta g^{(0)}_{ij} ) + \cdots \Bigg] ~.
	\label{eq Boundary term of the variation of a general action expanded in z}
    \end{aligned}
\end{equation}
The term of order $z_0^3$ inside the brackets contains the contractions ${P^{(3)}}^{zk}_{zi}$ and ${P^{(3)}}^{lk}_{li}$, which can be seen to vanish\footnote{Notice that in $D=4$ only $\Tr\,g^{(3)} = 0$, the other components (the off-trace part) being free, corresponding to the holographic stress-tensor in the dual CFT. In this case, ${P^{(3)}}^{lk}_{li} \neq 0$, thus inducing a constant term at the boundary which is standard.} provided $g^{(3)}_{ij} = 0$ following the reasoning of section \ref{subsec Vanishing of g3 in a general HCG}. So these divergences do not appear in the general theories that we are interested in.\footnote{The expansion of ${P^{(3)}}^{\mu\nu}_{\alpha\beta}$ also contains terms of the form $\nabla g^{(2)}$. However, notice that these are only present in the components ${P^{(3)}}^{z i}_{jk}$ or ${P^{(3)}}^{i j}_{z k}$, which are absent at this order ---albeit they might appear at higher orders in the expansion of the boundary term of the variation (\ref{eq Boundary term of the variation of a general action expanded in z}).} The contributions at order $z_0^2$ in \eqref{eq Boundary term of the variation of a general action expanded in z} also vanish, as can be seen by showing that, on-shell,
\begin{equation}
	{P^{(2)}}^{zk}_{zi} = {P^{(2)}}^{lk}_{li} = 0 ~.
\end{equation}
The arguments leading to this result are explained in appendix \ref{appendix Proof that the contributions of order $z^2$ in the boundary term of the variation vanish on-shell}. Finally, the term proportional to $z_0^4$ in \eqref{eq Boundary term of the variation of a general action expanded in z} vanishes as $z_0 \rightarrow 0$ in $D = 3$ and $D = 4$, even when multiplied by $\sqrt{-h}$. In $D = 5$, however, it produces a constant term which corresponds to a well posed variational problem with Dirichlet boundary conditions, as all the variations that it contains can be written in terms of $\delta g^{(0)}_{ij}$.

Gathering everything up, we see that the boundary term of the variation reads
\begin{equation}
    \delta I = - \int_{\partial M} d^{D-1}\! x \, \sqrt{-h} \, \Bigg[ C(L) \left( 2 \delta K^i_i + (h^{-1} \delta h)^i_j K^j_i \right) + z_0^4 \, \OO ( \delta g^{(0)}_{ij} ) + \cdots \Bigg] ~.
    \label{eq Boundary term of the variation of a general action expanded in z only divergent terms}
\end{equation}
By simple inspection, we see that the only divergent part up to $D = 5$ is the same as that of Einstein gravity, presented for example in \cite{Anastasiou:2019ldc}, multiplied by the constant $C(L)$ which depends on the couplings of the theory at hand. Therefore, we will assume that the divergences in this object can be regularized by adding to the original Lagrangian the usual boundary Kounterterms that are known to work for Einstein gravity, multiplied by the constant $C(L)$ as given in section \ref{section3}, and which we already know that are enough to cancel the divergences of the on-shell action in these dimensions.

\subsection{Explicit analysis in different dimensions}

We will now show how the Kounterterms presented in section \ref{section3} can be used to cancel the divergences in the boundary term that appears when varying the action on-shell \eqref{eq Boundary term of the variation of a general action expanded in z only divergent terms}. Although the following computations are carried out more generally in appendix D of \cite{Anastasiou:2019ldc}, here we will show the results more explicitly in up to 5 bulk dimensions.

\subsubsection{$D = 3$ dimensions}

Since $\sqrt{-h} \sim z^{-2}$ in this case, the non-vanishing terms of the variation \eqref{eq Boundary term of the variation of a general action expanded in z only divergent terms} are simply
\begin{equation}
	\delta I = - C(L) \int_{\partial M} d^2 x \, \sqrt{-h} \left( 2 \delta K^i_i + (h^{-1} \delta h)^i_j K^j_i \right) ~.
\end{equation}
In order to regularize this, we add the Kounterterm \eqref{eq General Kounterterms odd dimensions} particularized to $D = 3$, which reads
\begin{equation}
	I_\text{Kt} = C(L) \int_{\partial M} d^2 x \, \sqrt{-h} \, K ~.
\end{equation}
It is straightforward to compute its variation, finding
\begin{equation}
		\delta I_\text{Kt} = C(L) \int_{\partial M} d^2 x \, \sqrt{-h} \left( \delta K^i_i + \frac{1}{2} (h^{-1} \delta h)^i_i K \right) ~. 
\end{equation}
Then, adding this to $\delta I$ above we get the variation of the renormalized action in $D = 3$,
\begin{equation}
	\delta I_\text{reg} = \delta I + \delta I_\text{Kt} = C(L) \int_{\partial M} d^2 x \, \sqrt{-h} \left( \frac{1}{2} (h^{-1} \delta h)^i_i K - (h^{-1} \delta h)^i_j K^j_i - \delta K^i_i \right) ~.
\end{equation}
To the lowest orders in $z$, the extrinsic curvature behaves as
\begin{equation}
	K^i_j = \frac{1}{L} \delta^i_j + z^2 L S^i_j (g^{(0)}) + \cdots ~,
	\label{eq Extrinsic curvature in terms of Schouten}
\end{equation}
where $S^i_j (g^{(0)})$ is the Schouten tensor of the metric $g^{(0)}_{ij}$. So we see that the terms in the parenthesis start at order $z^2$, which is finite when multiplied by the determinant factor. Therefore, the Kounterterm cancels the divergences in the variation for this dimension of the spacetime, and also it allows the variation to be written only in terms of variations with respect to $g^{(0)}_{ij}$, thus leading to a well-posed variational problem.

\subsubsection{$D = 4$ dimensions}

The non-vanishing terms of the boundary variation of the action in 4 dimensions are
\begin{equation}
	\delta I = - C(L) \int_{\partial M} d^3 x \, \sqrt{-h} \left( 2 \delta K^i_i + (h^{-1} \delta h)^i_j K^j_i \right) ~.
	\label{eq Divergent boundary terms variation general theory 4 dimensions}
\end{equation}
These should be cancelled by the Kounterterm \eqref{eq General Kounterterms even dimensions} with $n = 2$, which is
\begin{equation}
    I_\text{Kt} = - L^2 C(L) \int_{\partial M} d^3 x \, \sqrt{-h} \, \delta^{i_1 i_2 i_3}_{j_1 j_2 j_3} K^{j_1}_{i_1} \left( \frac{1}{2} \RR^{j_2 j_3}_{i_2 i_3} - \frac{1}{3} K^{j_2}_{i_2} K^{j_3}_{i_3} \right) ~.
\end{equation}
Its variation $\delta I_\text{Kt}$ can be evaluated explicitly term by term, and the final result reads \cite{Anastasiou:2019ldc}
\begin{equation}
	\begin{aligned}
		\delta I_\text{Kt} & = C(L) \int_{\partial M} d^3 x \, \sqrt{-h} \left( 2 \delta K^i_i + (h^{-1} \delta h)^i_j K^j_i \right) \\
		& \hspace{1cm} + L^2 C(L) \int_{\partial M} d^3 x \, \sqrt{-h} \Big[ W^{il}_{jl} \left( (h^{-1} \delta h)^j_k K^k_i + 2 \delta K^j_i \right) \\
		& \hspace{5.5cm} - \delta^{ijk}_{lmn} K^{l}_{i} \tilde{\nabla}^{n} \tilde{\nabla}_{j} (h^{-1} \delta h)^{m}_{k} \Big] ~.
	\end{aligned}
	\label{eq Variation kounterterm 4 dimensions}
\end{equation}
The first term in this expression cancels exactly the divergent terms in the variation \eqref{eq Divergent boundary terms variation general theory 4 dimensions}. The second integral in $\delta I_\text{Kt}$ is finite if we assume asymptotic conformal flatness (which is natural, since any metric in three dimensions is Weyl-flat). Indeed, in this case the relevant components of the bulk Weyl tensor behave as $W^{il}_{jl} \sim z^{D-1} = z^3$, and since $\delta K^i_j \sim z^2$ due to \eqref{eq Extrinsic curvature in terms of Schouten}, only the first term in that parenthesis contributes. The last term of \eqref{eq Variation kounterterm 4 dimensions} can also be shown to vanish for this number of dimensions. Assuming that the boundary submanifold is infinite, we can integrate by parts without adding a boundary term,
\begin{equation}
	\sqrt{-h} \, \delta^{i_1 i_2 i_3}_{j_1 j_2 j_3} K^{j_1}_{i_1} \tilde{\nabla}^{j_3} \tilde{\nabla}_{i_2} (h^{-1} \delta h)^{j_2}_{i_3} \; \longrightarrow \; \sqrt{-h} \, \delta^{i_1 i_2 i_3}_{j_1 j_2 j_3} (h^{-1} \delta h)^{j_2}_{i_3} \tilde{\nabla}_{i_2} \tilde{\nabla}^{j_3} K^{j_1}_{i_2} ~.
\end{equation}
But since $\tilde{\nabla}_l K^{j_1}_{i_2} \sim z^2$ at least (the zeroth order of $K^i_j$ is proportional to $\delta^i_j$) and the indices of the covariant derivative are raised with the metric $h^{ij} = z^2 g^{ij}$, we have
\begin{equation}
	\sqrt{-h} \, \delta^{i_1 i_2 i_3}_{j_1 j_2 j_3} (h^{-1} \delta h)^{j_2}_{i_3} \tilde{\nabla}_{i_2} \tilde{\nabla}^{j_3} K^{j_1}_{i_2} \sim z ~,
\end{equation}
so this term vanishes as $z \rightarrow 0$. Therefore, the boundary term of the variation of the regularized action in $D = 4$ reads
\begin{equation}
	\delta I_\text{reg} = \delta I + \delta I_\text{Kt} = L^2 C(L) \int_{\partial M} d^3x \, \sqrt{-h} \, W^{il}_{jl} (h^{-1} \delta h)^j_k K^k_i ~,
\end{equation}
which is finite and can be written as depending only on variations of $g^{(0)}_{ij}$, thus leading to a well-posed variational problem with Dirichlet boundary conditions.

\subsubsection{$D = 5$ dimensions}\label{D=5 Variation}

In 5 dimensions the form of the divergent contribution is the same as before, with the difference that now the terms of order $z_0^4$ also contribute,
\begin{equation}
	\delta I = - \int_{\partial M} d^4x \, \sqrt{-h} \left[ C(L) \left( 2 \delta K^i_i + (h^{-1} \delta h)^i_j K^j_i \right) + z_0^4 \, \OO(\delta g^{(0)}_{ij}) \right] ~.
	\label{eq Divergent boundary terms variation general theory 5 dimensions}
\end{equation}
Notice that in this case only the first term inside the brackets produces divergences, and the second one results in a constant in the integrand, so it does not need to be cancelled. The boundary Kounterterm that should cancel these divergences is \eqref{eq General Kounterterms odd dimensions} particularized to $n = 2$,
\begin{equation}
    I_\text{Kt} = - \frac{L^2}{8} C(L) \int_{\partial M} d^4 x \, \sqrt{-h} \delta^{i_1 i_2 i_3}_{j_1 j_2 j_3} K^{j_1}_{i_1} \left( \RR^{j_2 j_3}_{i_2 i_3} - K^{j_2}_{i_2} K^{j_3}_{i_3} + \frac{1}{3 L^2} \delta^{j_2}_{i_2} \delta^{j_3}_{i_3} \right) ~.
\end{equation}
Obtaining the variation of this Kounterterm requires a rather involved computation, which again can be carried out following appendix D of \cite{Anastasiou:2019ldc}. The final result reads
\begin{equation}
    \delta I_\text{Kt} = C(L) \int_{\partial M} d^4 x \, \sqrt{-h} \left( 2 \delta K^i_i + (h^{-1} \delta h)^i_j K^j_i \right) + \delta I^{(W)} + \delta I^{(0)} + \delta I^{(\tilde{\nabla})} ~, 
\end{equation}
where the first term cancels exactly the divergent part of \eqref{eq Divergent boundary terms variation general theory 5 dimensions}, and we have defined
\begin{align}
	\delta I^{(W)} & = - \frac{L^2}{8} C(L) \int_{\partial M} d^4 x \, \sqrt{-h} \, \delta^{i_1 i_2 i_3}_{j_1 j_2 j_3} W^{j_2 j_3}_{i_2 i_3} \left( 2 \delta K^{j_1}_{i_1} + (h^{-1} \delta h)^{j_1}_k K^k_{i_1} \right) ~, \\[0.5em]
	\begin{split}
	    \delta I^{(0)} & = \frac{L^2}{16} C(L) \int_{\partial M} d^4 x \, \sqrt{-h} \, \delta^{i_1 i_2 i_3 i_4}_{j_1 j_2 j_3 j_4} \left( \RR^{j_3 j_4}_{i_3 i_4} - K^{j_3}_{i_3} K^{j_4}_{i_4} + \frac{1}{L^2} \delta^{j_3}_{i_3} \delta^{j_4}_{i_4} \right) \\
    	& \hspace{5cm} \times \left( (h^{-1} \delta h)^{j_1}_k ( K^k_{i_1} \delta^{j_2}_{i_2} - \delta^k_{i_1} K^{j_2}_{i_2} ) + 2 \delta^{j_1}_{i_1} \delta K^{j_2}_{i_2} \right) ~,
	\end{split} \\[0.4em]
	\delta I^{(\tilde{\nabla})} & = \frac{L^2}{4} C(L) \int_{\partial M} d^4 x \, \sqrt{-h} \, \delta^{i_1 i_2 i_3}_{j_1 j_2 j_3} (h^{-1} \delta h)^{j_2}_{i_3} \tilde{\nabla}_{i_2} \tilde{\nabla}^{j_3} K^{j_1}_{i_1} ~.
\end{align}
Then, the variation of the total regularized action is
\begin{equation}
	\delta I_\text{reg} = \delta I^{(W)} + \delta I^{(0)} + \delta I^{(\tilde{\nabla})} + \delta I^{(z_0^4)},
	\label{eq Variation regularized action 5 dimensions symbolical}
\end{equation}
where $\delta I^{(z_0^4)}$ corresponds to the terms of order $z_0^4$ in $\delta I$ that produce a constant in the integrand, and whose particular form depends on the theory. In order to show that the variation \eqref{eq Variation regularized action 5 dimensions symbolical} of the regularized action is finite, we should count the powers of $z$ appearing in each of the terms.
\begin{itemize}
\item $\delta I^{(W)}$ can be rewritten by expanding the sum in the indices of the antisymmetric $\delta$, and using $W^{ij}_{ij} = 0$, which follows from $W^{\mu\nu}_{\mu\nu} = W^{\mu i}_{\mu i} = 0$. We find:
\begin{equation}
	\delta I^{(W)} = \frac{L^2}{2} C(L) \int_{\partial M} d^4 x \, \sqrt{-h} \, W^{il}_{jl} \left( 2 \delta K^j_i + (h^{-1} \delta h)^j_k K^k_i \right) ~.
\end{equation}
But now recall that $\delta K^i_j \sim z^2$, and under the assumption of asymptotic conformal flatness, $W^{il}_{jl} \sim z^{D-1} = z^4$. Then, since $\sqrt{-h} \sim z^{-4}$, the term with $\delta K^j_i$ in the parenthesis does not contribute, and we can write simply
\begin{equation}
	\delta I^{(W)} = \frac{L^2}{2} C(L) \int_{\partial M} d^4 x \, \sqrt{-h} \, W^{il}_{jl} (h^{-1} \delta h)^j_k K^k_i ~.
\end{equation}
\item The first parenthesis in $\delta I^{(0)}$ can be rewritten in terms of the Weyl tensor of the bulk metric, using the Gauss-Codazzi equation
\begin{equation}
	R^{ij}_{kl} = \RR^{ij}_{kl} - 2 K^i_{[k} K^j_{l]} ~,
\end{equation}
and the definition of the Weyl tensor, which to the lowest order in $z$ yields:
\begin{equation}
	W^{ij}_{kl} = R^{ij}_{kl} + \frac{2}{L^2} \delta^i_{[k} \delta^j_{l]} ~.
\end{equation}
These two expressions can be combined to find
\begin{equation}
	\RR^{ij}_{kl} = W^{ij}_{kl} + 2 K^i_{[k} K^j_{l]} - \frac{2}{L^2} \delta^i_{[k} \delta^j_{l]} ~.
\end{equation}
The first parenthesis in $\delta I^{(0)}$ (taking into account the prefactor $\delta^{i_1 i_2 i_3 i_4}_{j_1 j_2 j_3 j_4}$) can now be rewritten as 
\begin{equation}
	\RR^{j_3 j_4}_{i_3 i_4} - K^{j_3}_{i_3} K^{j_4}_{i_4} + \frac{1}{L^2} \delta^{j_3}_{i_3} \delta^{j_4}_{i_4} = W^{j_3 j_4}_{i_3 i_4} + K^{j_3}_{i_3} K^{j_4}_{i_4} - \frac{1}{L^2} \delta^{j_3}_{i_3} \delta^{j_4}_{i_4} ~,
\end{equation}
and since $W^{i j}_{k l} \sim z^4$ and $K^i_j = \delta^i_j / L + \OO(z^2)$, we see that, to the lowest order,
\begin{equation}
	\RR^{j_3 j_4}_{i_3 i_4} - K^{j_3}_{i_3} K^{j_4}_{i_4} + \frac{1}{L^2} \delta^{j_3}_{i_3} \delta^{j_4}_{i_4} \sim z^2 ~.
\end{equation}
In the second parenthesis in $\delta I^{(0)}$, we have
\begin{equation}
	K^k_{i_1} \delta^{j_2}_{i_2} - \delta^k_{i_1} K^{j_2}_{i_2} \sim z^2, \qquad \delta K^{j_2}_{i_2} \sim z^2 ~.
\end{equation}
Therefore, the whole integrand starts at order $z^4$, and when integrated with $d^4 x \, \sqrt{-h}$ it produces a term that is constant and thus non-divergent when $z \rightarrow 0$.
\item If we do a naive power counting in the term $\delta I^{(\tilde{\nabla})}$, we could find that it produces a constant at the boundary, $z \rightarrow 0$. Indeed, as we have seen in $D = 4$, $\tilde{\nabla}_l K^{j_1}_{i_2} \sim z^2$. Therefore, we might be tempted to think that
\begin{equation}
    \sqrt{-h} \, \delta^{i_1 i_2 i_3}_{j_1 j_2 j_3} (h^{-1} \delta h)^{j_2}_{i_3} \tilde{\nabla}_{i_2} \tilde{\nabla}^{j_3} K^{j_1}_{i_1} \sim 1 ~.
\end{equation}
However, we will now show that this expression vanishes if we impose the boundary to be conformally flat. Indeed, since we can expand $K^i_j$ in terms of the Schouten tensor as in \eqref{eq Extrinsic curvature in terms of Schouten}, to leading order in $z$ we can write
\begin{equation}
    \begin{aligned}
    	\tilde{\nabla}_{i_2} \tilde{\nabla}^{j_3} K^{j_1}_{i_1} & = z^2 L \tilde{\nabla}_{i_2} \tilde{\nabla}^{j_3} {S^{(0)}}^{j_1}_{i_1} + \cdots = z^4 L \tilde{\nabla}^{(0)}_{i_2} \tilde{\nabla}^{(0) j_3} {S^{(0)}}^{j_1}_{i_1} + \cdots \\[0.6em]
		& = z^4 \frac{L}{2} \tilde{\nabla}^{(0)}_{i_2} \left( \tilde{\nabla}^{(0) j_3} {S^{(0)}}^{j_1}_{i_1} - \tilde{\nabla}^{(0) j_1} {S^{(0)}}^{j_3}_{i_1} \right) + \cdots,
    \end{aligned}
\end{equation}
where $\tilde{\nabla}^{(0)}_i$ is the covariant derivative compatible with the boundary metric $g^{(0)}_{ij}$, and the indices of the objects with superscript $(0)$ are raised using the inverse metric $g^{(0)}_{ij}$ (thus the extra explicit $z^2$ factor in the second step). In the last step, we used the fact that this object is contracted with a generalized Kronecker $\delta$, so we antisymmetrized it explicitly in the indices $j_1$ and $j_3$. This allows us to use the definition of the Cotton tensor,
\begin{equation}
	C^{(0)}_{ijk} = \tilde{\nabla}^{(0)}_k S^{(0)}_{ij} - \tilde{\nabla}^{(0)}_j S^{(0)}_{ik} ~,
\end{equation}
in order to rewrite the above expression as
\begin{equation}
	 \tilde{\nabla}_{i_2} \tilde{\nabla}^{j_3} K^{j_1}_{i_1} = z^4 \frac{L}{2} \tilde{\nabla}^{(0)}_{i_2} {C^{(0)}}^{j_3 j_1}_{i_1} + \cdots ~.
\end{equation}
But the Cotton tensor of $g^{(0)}_{ij}$ is related to its Weyl tensor \cite{Anastasiou:2020zwc}
\begin{equation}
	{C^{(0)}}_j^{kl} = \frac{1}{D-4} \tilde{\nabla}^{(0) i} {W^{(0)}}^{kl}_{ij} ~,
\end{equation}
which is zero if we impose asymptotic conformal flatness, ${W^{(0)}}^{ij}_{kl} = 0$. Therefore, the term of order $z^4$ in $\tilde{\nabla}_{i_2} \tilde{\nabla}^{j_3} K^{j_1}_{i_1}$ vanishes, and $\tilde{\nabla}_{i_2} \tilde{\nabla}^{j_3} K^{j_1}_{i_1} \sim z^6$, which means that the total integrand in $\delta I^{(\tilde{\nabla})}$ is zero in $D = 5$ dimensions, since
\begin{equation}
	\sqrt{-h} \, \delta^{i_1 i_2 i_3}_{j_1 j_2 j_3} (h^{-1} \delta h)^{j_2}_{i_3} \tilde{\nabla}_{i_2} \tilde{\nabla}^{j_3} K^{j_1}_{i_1} \sim z^2
\end{equation}
vanishes at the boundary.
\end{itemize}
Gathering everything up, we find that the boundary term of the variation of the regularized action in $D = 5$ is
\begin{equation}
	\delta I_\text{reg} = \int_{\partial M} d^4 x \, \sqrt{-h} \left[ \frac{L^2}{2} C(L) 
	W^{il}_{jl} (h^{-1} \delta h)^j_k K^k_i + z_0^4 \, \OO(\delta g^{(0)}_{ij}) \right] ~,
\end{equation}
which yields a well-posed Dirichlet variational problem. The last term between brackets contains the terms of order $z_0^4$ appearing in the original variation \eqref{eq Divergent boundary terms variation general theory 5 dimensions}, as well as those coming from $\delta I^{(0)}$.

\section{Discussion}
\label{section6}

In this work, we have implemented a universal renormalization procedure applicable to arbitrary higher curvature gravity theories evaluated on AlAdS manifolds of $D\leq5$ dimensions.\footnote{In the case of $D=5$, the extra condition of Asymptotically Conformal Flatness \cite{Anastasiou:2019ldc} is required, which assumes the manifold to have a conformally flat boundary. This condition is needed to guarantee that $g_{(2)}$ has the universal form of \eqref{eq Coefficient g2 general form}, and for the variational principle to be well-posed, as discussed in section \ref{D=5 Variation}.} This method uses the extrinsic boundary counterterms of \cite{Olea:2005gb,Olea:2006vd}, but with a theory-dependent coupling constant, as given in equations \eqref{eq Coupling Kounterterm general even dimensions} and \eqref{eq Coupling Kounterterm general odd dimensions}. In order to show the universality of the method, we decompose the equations of motion of an arbitrary HCG into their radial and tangential components (with respect to the holographic Poincare coordinate), and we are able to argue, on general grounds ---by symmetry arguments--- that the odd coefficients of the FG expansion of the bulk metric, $g^{(1)}_{ij}$ and $g^{(3)}_{ij}$, are zero. Furthermore, by virtue of the PBH transformation relations \cite{Imbimbo:1999bj}, it can be argued that $g^{(2)}_{ij}$ is constrained to have the universal form \eqref{eq Coefficient g2 general form}. Then, considering these general features of the FG expansion, we verify the cancellation of divergences (section \ref{section4}) and the well-posedness of the variational principle (section \ref{section5}), keeping the expansion terms up to the normalizable order.

The argument fails for particular theories (discussed in section \ref{subsec Vanishing of g1 in a general HCG} and appendix \ref{appendix Conditions to leave g1 undetermined in cubic theories of gravity}), for which the equations of motion do not constrain the form of the $g^{(1)}_{ij}$ and/or the $g^{(3)}_{ij}$ coefficients. Albeit these theories correspond to zero measure submanifolds in theory space (parameterized by the couplings of the higher curvature terms), they are interesting on their own, as they include theories displaying degenerate AdS vacua ---those that leave $g^{(1)}$ undetermined--- and modified AdS asymptotics (as discussed in section \ref{subsec Vanishing of g1 in a general HCG}).

It is natural to think about applying this procedure to obtain the finite asymptotic charges for black hole solutions in generic HCGs. Also, in the context of the AdS/CFT correspondence, one could use the method for renormalizing holographic entanglement entropy, as in \cite{Anastasiou:2017xjr,Anastasiou:2018rla,Anastasiou:2019ldc,Anastasiou:2021swo,Anastasiou:2021jcv}. Finally, one could use the conditions obtained in \eqref{g1 condition}, which are required in order for $g^{(1)}_{ij}$ not to be fixed by the equations of motion, to seek for new theories with degenerate AdS vacua and/or modified asymptotic behaviour. Certainly, many interesting avenues of exploration are possible using this universal renormalization approach.

\acknowledgments

IJA thanks Giorgos Anastasiou and Rodrigo Olea for interesting discussions.
The work of IJA is funded by Agencia Nacional de Investigación y Desarrollo (ANID), REC Convocatoria Nacional Subvenci\'on a Instalaci\'on en la Academia Convocatoria A\~no 2020, Folio PAI77200097.
The work of JDE, AR, DVR and AVL is supported by MINECO FPA2017-84436-P, Xunta de Galicia ED431C 2017/07, Xunta de Galicia (Centro singular de investigaci\'on de Galicia accreditation 2019-2022), the European Union (European Regional Development Fund -- ERDF), the ``Mar\'\i a de Maeztu'' Units of Excellence MDM-2016-0692, and the Spanish Research State Agency.
ARS is supported by the Spanish MECD fellowship FPU18/03719.
DVR is supported by Xunta de Galicia under the grant ED481A-2019/115.
AVL is supported by the Spanish MECD fellowship FPU16/06675.
AVL is pleased to thank the Instituto de F\'\i sica Te\'orica (IFT, Madrid) where part of this work was done, for their warm hospitality.

\appendix

\section{Conditions to leave $g^{(1)}_{ij}$ undetermined in cubic theories}
\label{appendix Conditions to leave g1 undetermined in cubic theories of gravity}

In the end of section \ref{subsec Vanishing of g1 in a general HCG} we mentioned some quadratic theories of gravity that do not impose $g^{(1)}_{ij} = 0$, but leave it free to be determined as a boundary condition. Here, we will do the same for theories constructed with cubic contractions of the curvature tensors. Let us consider the most general theory constructed from the Einstein-Hilbert action supplemented by all possible terms that are cubic in the curvature,
\begin{equation}
	\begin{aligned}
		I = & \int_M d^Dx \sqrt{-g} \bigg( \frac{R-2\Lambda_0}{\kappa} + \lambda_1 R^{\mu\nu\rho\sigma} \tensor{R}{_\mu^\gamma_\rho^\delta} R_{\nu\gamma\sigma\delta} + \lambda_2 \tensor{R}{_\mu_\nu^\gamma^\delta} R^{\mu\nu\rho\sigma} R_{\rho\sigma\gamma\delta} \\[0.6em]
		& \hspace{2cm} + \lambda_3 R^{\mu\nu} \tensor{R}{_\mu^\rho^\sigma^\gamma} \tensor{R}{_\nu_\rho_\sigma_\gamma} + \lambda_4 R R_{\mu\nu\rho\sigma} R^{\mu\nu\rho\sigma} + \lambda_5 R^{\mu\nu} R^{\rho\sigma} R_{\mu\nu\rho\sigma} \\[0.6em]
		& \hspace{4cm} + \lambda_6 R^{\mu\nu} R_{\nu\rho} \tensor{R}{_\mu^\rho} + \lambda_7 R_{\mu\nu} R^{\mu\nu} R + \lambda_8 R^3 \bigg) ~,
	\end{aligned}
	\label{eq Lagrangian cubic curvature gravity general}
\end{equation}
where the coupling constants $\lambda_i$ are arbitrary for the moment. The value of the constant $C(L)$ that multiplies the Kounterterms can be computed using equation \eqref{eq Constant C(L) general},
\begin{equation}
	\begin{aligned}
		C(L) = \frac{1}{\kappa} + \frac{3 (D-1)}{L^4} & \bigg( \frac{D-2}{D-1} \lambda_1 + \frac{4}{D-1} \lambda_2 + 2 \lambda_3 + 2 D \lambda_4 + (D-1) \lambda_5 \\[0.6em]
		& \hspace{1cm} + (D-1) \lambda_6 + D (D-1)  \lambda_7 + D^2 (D-1) \lambda_8 \bigg) ~. 
	\end{aligned}
\end{equation}
The equations of motion at order $z$ \eqref{eq Projected equation nn general theory order z1}, \eqref{eq Projected equation pp general theory order z1} and \eqref{eq Projected equation np general theory order z1} are determined by the constants $a^{(1)}(L)$ and $b^{(1)}(L)$, which in this case are given by
\begin{align}
    \begin{split}
        a^{(1)}(L) & = \frac{1}{\kappa} + \frac{1}{L^4} \bigg[ 6 (D-3) \lambda_1 + 36 \lambda_2 + 2 (7D-9) \lambda_3 + 10 D (D-1) \lambda_4 + (5D^2-13D+9)\lambda_5 \nonumber \\[0.6em]
        & \hspace{2.3cm} + (6D^2-15D+9)\lambda_6 + D (4D^2-9D+5) \lambda_7 + 3 D^2 (D-1)^2 \lambda_8 \bigg] ~,
    \end{split} \\[0.5em]
    \begin{split}
        b^{(1)}(L) & = - \frac{1}{\kappa} + \frac{1}{L^4} \bigg[ 6 \lambda_1 + 12 \lambda_2 + 2 (D+5) \lambda_3 - 2 (D^2-17D+16) \lambda_4 - (D^2-15D+17) \lambda_5 \nonumber \\[0.6em]
        & \hspace{2cm} + 9 (D-1) \lambda_6 - (2D^3-23D^2+37D-16) \lambda_7 - 3 D (D-1)^2 (D-8) \lambda_8 \bigg] ~.
    \end{split}
\end{align}
As explained in section \ref{subsec Vanishing of g1 in a general HCG}, the equations of motion imply $g^{(1)}_{ij} = 0$, unless the conditions discussed after \eqref{vanishing-g1} are met. Thus, only for both $a^{(1)}(L) = b^{(1)}(L) = 0$, $g^{(1)}_{ij}$ is fully unconstrained by the equations of motion, which happens only in a zero measure region of the space of parameters $\lambda_i$. Besides, as what was found for quadratic theories of gravity, the conditions $a^{(1)}(L) = b^{(1)}(L) = 0$ end up implying that the corresponding cubic theory has degenerate AdS vacua.

In the case of cubic curvature gravity, let us quote two examples of usually well-behaved theories which have the above properties:
\begin{itemize}
\item Cubic Lovelock theory in general dimensions \cite{Lovelock:1971yv}, $D \geq 7$. This corresponds to setting in \eqref{eq Lagrangian cubic curvature gravity general} the values of the couplings to be
\begin{gather}
    \nonumber \lambda_1 = -8 \mu ~, \quad \lambda_2 = 4 \mu ~, \quad \lambda_3 = -24 \mu ~, \quad \lambda_4 = 3 \mu ~, \quad \lambda_5 = 24 \mu ~, \\[0.6em]
	\lambda_6 = 16 \mu ~, \quad \lambda_7 = -12 \mu ~, \quad \lambda_8 = \mu ~.
    \end{gather}
The particular value of the coupling $\mu$ at which $a^{(1)}(L) = b^{(1)}(L) = 0$ corresponds to
\begin{equation}
	\mu = - \frac{L^4}{3 (D-3) (D-4) (D-5) (D-6) \kappa} ~,
\end{equation}
which corresponds to the critical value \cite{Camanho:2009hu}.
\item Einsteinian Cubic Gravity \cite{Bueno:2016xff}. We could consider the Lagrangian $(R - 2 \Lambda_0) / \kappa + \mu_\mathcal{P} \mathcal{P}$, which amounts to
\begin{equation}
    \lambda_1 = 12 \mu_\mathcal{P} ~, \qquad \lambda_2 = \mu_\mathcal{P} ~, \qquad \lambda_5 = -12 \mu_\mathcal{P} ~, \qquad \lambda_6 = 8 \mu_\mathcal{P} ~.
\end{equation}    
while the remaining couplings vanish.
The coefficient $g^{(1)}_{ij}$ becomes undetermined at the critical value of the coupling
\begin{equation}
    \mu_\mathcal{P} = \frac{L^4}{12 (D-3) (D-6) \kappa} ~,
    \label{eq Critical coupling Einsteinian cubic gravity}
\end{equation}
which corresponds to the critical value found in \cite{Bueno:2018xqc} when studying the AdS vacua of the theory in 4 dimensions.
    
One could also consider the Lagrangian density $\mathcal{C}$, defined in \cite{Hennigar:2017ego}. In particular, the combination $\mathcal{P} - 8 \mathcal{C}$ in $D = 4$, introduced in \cite{Arciniega:2018tnn,Arciniega:2018fxj} due to its interesting cosmological properties, also leaves the coefficient $g^{(1)}_{ij}$ undetermined for the value of the coupling \eqref{eq Critical coupling Einsteinian cubic gravity}. This makes sense, as it is known that $\mathcal{C}$ does not modify the AdS vacuum in four dimensions. 
\end{itemize}
It would be interesting to explore whether the $a^{(1)}(L) = b^{(1)}(L) = 0$ condition can be used as a tool to look for new theories in higher dimensions and of higher order in the curvature, with analogous behaviour to the examples discussed here.

\section{Conditions to leave $g^{(3)}_{ij}$ undetermined in quadratic and cubic theories}
\label{appendix Constants in the projected equations of motion at order z3 for general quadratic and cubic theories}

In this appendix we give the values of the constants $a^{(3)}(L)$ and $b^{(3)}(L)$ introduced in section \ref{subsec Vanishing of g3 in a general HCG} for different families of theories. These determine the projected equations of motion at third order in $z$, \eqref{eq Projected equation nn general theory order z3} and \eqref{eq Projected equation pp general theory order z3}. First let us consider the general quadratic gravity action \eqref{eq Lagrangian quadratic curvature gravity general}. These constants are expressions of the AdS radius $L$ and the coupling constants in the Lagrangian, and read:
\begin{eqnarray}
    a^{(3)}(L) & = & \frac{3}{\kappa} + \frac{1}{L^2} \bigg[ -3 (5D-14) \alpha_1 - 6 D (D-1) \alpha_2 - 6 (D-3) (D-4) \alpha_3 \bigg] ~, \nonumber \\ [0.6em]
    b^{(3)}(L) & = & - \frac{3}{\kappa} + \frac{1}{L^2} \bigg[ -3 (D-6) \alpha_1 + 6 (D^2-9D+24) \alpha_2 + 6 (D-3) (D-4) \alpha_3 \bigg] ~. \nonumber
\end{eqnarray}
For the general theory containing cubic contractions of the curvature tensors \eqref{eq Lagrangian cubic curvature gravity general}, they take the values:
\begin{eqnarray}
    a^{(3)}(L) & = & \frac{3}{\kappa} + \frac{1}{L^4} \bigg[ 36 \lambda_1 + 36 (4D-17) \lambda_2 + 6 (4D^2-13D-9) \lambda_3 + 6 D (D-1) (4D-15) \lambda_4 \nonumber \\[0.4em]
    & &\hspace{2.3cm} + 9 (3D^2-13D+13) \lambda_5 + 9 (D-1) (4D-13) \lambda_6 \\[0.4em]
    & &\hspace{2.3cm} + 9 D (D-1) (2D-5) \lambda_7 + 9 D^2 (D-1)^2 \lambda_8 \bigg] ~, \nonumber \\[0.6em]
    b^{(3)}(L) & = & - \frac{3}{\kappa} + \frac{1}{L^4} \bigg[ 18 (D-4) \lambda_1 + 36 \lambda_2 + 30 (D-3) \lambda_3 - 6 (D^2-33D+96) \lambda_4 \nonumber \\[0.4em]
    & & \hspace{2.3cm} + 9 (D^2-D-9) \lambda_5 + 9 (D-1) (2D-7) \lambda_6 \\[0.4em]
    & & \hspace{2.3cm} + 9 (D-1) (9D-32) \lambda_7 - 9 D (D-1) (D^2-17D+48) \lambda_8 \bigg] ~. \nonumber
\end{eqnarray}
Studying the particular points where they vanish might lead us to theories of gravity whose dynamics differs from that of Einstein gravity.

\section{Proof that the boundary term of $\delta I$ vanishes on-shell at $\OO(z^2)$}
\label{appendix Proof that the contributions of order $z^2$ in the boundary term of the variation vanish on-shell}

In this appendix we will show that the terms of order $z_0^2$ in \eqref{eq Boundary term of the variation of a general action expanded in z} vanish for a general HCG. This contribution contains the quantity
\begin{equation}
    - 2 (2D-7) {P^{(2)}}^{zk}_{zi} + 4 {P^{(2)}}^{lk}_{li} ~,
\end{equation}
and thus, in order to prove that it vanishes it is enough to show that the two contractions of ${P^{(2)}}^{\mu\nu}_{\rho\sigma}$ appearing here vanish when evaluated on-shell. The tensor $P^{\mu\nu}_{\rho\sigma}$ is defined as the derivative of the Lagrangian $\LL (R^{\alpha\beta}_{\gamma\delta})$ with respect to the Riemann tensor. Then, its components will be given by contractions of the Riemann with four free indices that fulfill the symmetries of the Riemann tensor itself. Since we are interested in the form of the terms at order $z^2$ in this tensor, ${P^{(2)}}^{\mu\nu}_{\rho\sigma}$, we need to study the components of the Riemann up to this order, which are given once the equations of motion are fulfilled by
\begin{equation}
	\begin{aligned}
		R^{ij}_{kl} & = - \frac{2}{L^2} \delta^{[i}_k \delta^{j]}_l + z^2 \left( \frac{4}{L^2} \delta^{[i}_{[k} {g^{(2)}}^{j]}_{l]} + \RR^{i j}_{k l} \right) + \OO(z^4) ~, \\[0.6em]
		R^{z i}_{z j} & = - \frac{1}{L^2} \delta^i_j + \OO(z^4) ~, \\[0.6em]
		R^{z i}_{j k} & = \OO (z^3) ~.
	\end{aligned}
	\label{eq Components Riemann up to order z2}
\end{equation}
As said, also contractions of the curvature can contribute to $P^{\mu\nu}_{\rho\sigma}$ in a general theory. In particular, it is enough to consider $R^{ik}_{jk}$ and $R^{z i}_{z i}$. If we impose that the equations of motion are fulfilled, $g^{(2)}_{ij}$ is given by \eqref{eq Coefficients FG expansion Einstein gravity} and thus the form of these contractions is found to be
\begin{equation}
	\begin{aligned}
		R^{i k}_{j k} & = - \frac{D-2}{L^2} \delta^i_j + \OO(z^4) ~, \\[0.6em]
		R^{z i}_{z i} & = - \frac{D-1}{L^2} + \OO (z^4) ~.
	\end{aligned}
\end{equation}
Thereby we see that the uncontracted components $R^{ij}_{kl}$ are the only ones that can contribute to $P^{\mu\nu}_{\rho\sigma}$ at order $z^2$ on-shell. Therefore, we can write
\begin{equation}
	\begin{aligned}
		{P^{(2)}}^{ij}_{kl} & = C^{(2)} \left( \frac{4}{L^2} \delta^{[i}_{[k} {g^{(2)}}^{j]}_{l]} + \RR^{i j}_{k l} \right), \\[0.6em]
		{P^{(2)}}^{zi}_{zj} & = {P^{(2)}}^{zi}_{jk} = 0,
	\end{aligned}
\end{equation}
where $C^{(2)}$ is a constant depending upon the parameters of the particular theory that we consider. However, these expressions are enough to see that
\begin{equation}
    {P^{(2)}}^{zk}_{zi} = {P^{(2)}}^{lk}_{li} = 0 ~,
\end{equation}
and thus the terms at order $z^2$ in the boundary term of the variation of the action \eqref{eq Boundary term of the variation of a general action expanded in z} are zero on-shell for any theory of gravity.

\bibliography{GeneralHOGs.bib}
\bibliographystyle{JHEP}
\end{document}